\documentclass[10pt,journal,epsfig,twoside]{IEEEtran}
\usepackage{amsmath}
\usepackage{graphicx} 
\usepackage{epstopdf}
\usepackage{amssymb}
\usepackage{amsfonts}
\usepackage{amsthm}
\usepackage{cite}
\usepackage{bm,comment}
\usepackage{algorithm}
\usepackage{algpseudocode}

\usepackage{subeqnarray}
\usepackage{subfigure}
\usepackage{multicol}
\usepackage{multirow}
\usepackage{diagbox}
\usepackage{slashbox}
\usepackage{stfloats}
\usepackage{float}
\usepackage{color} 
\usepackage{cases}

\floatname{algorithm}{Algorithm}
\renewcommand{\algorithmicrequire}{\textbf{Given:}} 
\renewcommand{\algorithmicensure}{\textbf{Return:}}

\newtheorem{proposition}{Proposition}
\newtheorem{remark}{Remark}

\newtheorem{lemma}{Lemma}

\graphicspath{{figs/}}

\begin{document}
\IEEEoverridecommandlockouts
\title{Codebook Design and Performance Analysis for Wideband Beamforming in Terahertz Communications}
\author{Boyu Ning, \IEEEmembership{Member, IEEE}, Weidong Mei, \IEEEmembership{Member, IEEE}, Lipeng Zhu, \IEEEmembership{Member, IEEE}, Zhi Chen, \IEEEmembership{Senior Member, IEEE},  and Rui Zhang \IEEEmembership{Fellow, IEEE}, 

\thanks{Part of this work has been presented by the IEEE ICC, 2023\cite{maxmin}.}
\thanks{Boyu Ning, Weidong Mei and Zhi Chen are with National Key Laboratory of Science and Technology on Communications, University of Electronic Science and Technology of China, Chengdu 611731, China (e-mails: boydning@outlook.com; wmei@uestc.edu.cn; chenzhi@uestc.edu.cn).}
 \thanks{Lipeng Zhu is with the Department of Electrical and Computer Engineering, National University of Singapore, Singapore 117583 (e-mail: zhulp@nus.edu.sg).}
\thanks{R. Zhang is with School of Science and Engineering, Shenzhen Research Institute of Big Data, The Chinese University of Hong Kong, Shenzhen, Guangdong 518172, China (e-mail: rzhang@cuhk.edu.cn). He is also with the Department of Electrical and Computer Engineering, National University of Singapore, Singapore 117583 (e-mail: elezhang@nus.edu.sg). }
}
\maketitle

\maketitle

\begin{abstract}
The codebook-based analog beamforming is appealing for future terahertz (THz) communications since it can generate high-gain directional beams with low-cost phase shifters via low-complexity beam training. However, conventional beamforming codebook design based on array response vectors for narrowband communications may suffer from severe performance loss in wideband systems due to the ``beam squint" effect over frequency. To tackle this issue, we propose in this paper a new codebook design method for analog beamforming in wideband THz systems. In particular, to characterize the analog beamforming performance in wideband systems, we propose a new metric termed wideband beam gain, which is given by the minimum beamforming gain over the entire frequency band given a target angle. Based on this metric, a wideband analog beamforming codebook design problem is formulated  for optimally balancing the beamforming gains in both the spatial and frequency domains, and the performance loss of conventional narrowband beamforming in wideband systems is analyzed. To solve the new wideband beamforming codebook design problem, we divide the spatial domain into orthogonal angular zones each associated with one beam,  thereby decoupling the codebook design into a zone division sub-problem  and a set of beamforming optimization sub-problems each for one zone. For the zone division sub-problem, we propose a bisection method to obtain the optimal boundaries for separating adjacent zones. While for each of the per-zone-based beamforming optimization sub-problems, we further propose an efficient augmented Lagrange method (ALM) to solve it.  Numerical results demonstrate the performance  superiority of our proposed codebook design for wideband analog beamforming to the narrowband beamforming codebook and also validate our performance analysis.
\end{abstract}

\section{Introduction}
Terahertz (THz) communication, over a wide frequency band from $0.1$ THz to $10$ THz,  is a promising technology to solve the spectrum congestion problem in today's fifth-generation (5G) wireless communication systems\cite{arv}. In particular, compared with the millimeter-wave (mmWave) communication with several gigahertz (GHz) bandwidth (e.g., 2 GHz in IEEE 802.11ay\cite{802}), THz communication offers  more than ten times wider bandwidth and thus are expected to enable  drastically  higher data rate and lower latency for the future sixth-generation (6G) wireless systems\cite{sur1,cxw,irs6g}. However,  THz communication systems also face critical challenges in practical implementation, such as more  severe propagation loss over THz frequency band as compared to mmWave communications\cite{sur2}. To compensate for such high signal  loss, one potential solution is by equipping the THz base stations (BSs)/terminals with more smaller-size antenna elements,  thereby generating higher  directional beamforming gains than their mmWave counterparts \cite{hsa}.

The implementation of a massive antenna array for THz systems is practically cost-ineffective with a fully digital architecture due to the prohibitively high hardware costs and power consumption\cite{hangou}. This is because each antenna element is connected to a separate radio frequency (RF) chain, similar to traditional small-scale antenna arrays\cite{hb1}. As such, hybrid analog and digital array architectures are more practically feasible for THz systems, by striking a balance between communication performance and hardware costs\cite{hb3,hb4}. There are generally  three types of hybrid array architectures that have been considered extensively, i.e., the fully-connected architecture, the partially-connected architecture, and the dynamically-connected architecture. In practice, the fully-connected architecture needs the antenna-crossing transmission lines, while the dynamically-connected architecture needs an antenna-switching network, both of which incur practically undesired  wiring  and heat dissipation problems\cite{cross}. Thus, the partially-connected array architecture, usually referred to as the array of sub-arrays (AoSA)\cite{AoSA}, is regarded as an appealing low-cost solution for THz hardware implementation, where  the antenna elements are divided into multiple groups, each sharing a common RF chain\cite{chan}. The antenna elements in the same group form a sub-array and are connected  to individual  phase shifters\cite{AoSA2}, which enable  the {\it analog beamforming} by jointly tuning their phase shifts. 

However, different from the fully digital architecture where different beamforming vectors can be applied over different frequency bands,  analog beamforming is frequency-flat and cannot be tuned flexibly for different frequency bands based on their different channel realizations (e.g., in frequency-selective wideband channels) \cite{fs1,fs2,fs3}.  In conventional narrowband communications,  analog beamforming can be set based on the array response vector corresponding to the carrier frequency to align with the user's direction over the given  frequency band. However, for wideband communications in THz bands, due to the much larger signal bandwidth than that in sub-6 GHz/mmWave frequency bands, the signal components at different frequencies experience more distinct phase shifts, which results in more significantly varying gains of analog beamforming over the entire frequency band, also referred to as the ``beam squint'' effect\cite{bsquint,bsquint2}.

Some prior works have analyzed the communication performance loss due to beam squint in hybrid beamforming systems\cite{whb1,whb2}, but without any solutions given to deal with it.  The authors in \cite{whb3} employed multiple RF chains to create frequency-dependent beamforming vectors to alleviate the beam squint effect, which, however, induces  higher hardware cost.  The authors in \cite{split2} proposed to add true-time-delayers (TTDs) to the phase shifters  to compensate for the beam squint effect, which, however, needs to modify  the current transceiver hardware  and thus faces difficulty in practical implementation\cite{ttd5}. The authors in \cite{squint} proposed to increase the codebook size to reduce  the beam squint resulted performance loss. However, this method only increases the accuracy of beam alignment, which cannot further improve performance if the beam is already well aligned with the user direction.  Considering wideband performance, the authors in \cite{hybridsq} proposed a hybrid precoding design that aims to maximize the sum-rate across the frequency band for a user. This scheme requires perfect channel state information (CSI) and an adaptive OFDM scheme \cite{ofdm}, which are seldom achieved in practical systems.

To practically address the beam squint issue in wideband THz analog beamforming, our previous work \cite{maxmin} introduced a max-min beamformer designed to maximize the minimum beam gain across the entire frequency band, under the assumption of known CSI. In this paper, we advance our research by developing a codebook-based scheme that can be implemented with low-complexity beam training, eliminating the need for explicit CSI estimation. Specifically, codebook-based beamforming eliminates the necessity for channel estimation and beamforming optimization processes. This method involves scanning through predefined beams and selecting the most appropriate one, a process referred to as beam training. To the best of our knowledge, there have been some narrowband codebook designs for analog beamforming\cite{xiaotvt,wide3, hb2}, while the wideband analog beamforming codebook design is still lacking. The main contributions of this paper are summarized as follows:
\begin{itemize} 
\item First, to characterize the impact of beam squint on the performance of wideband analog beamforming, we propose a new performance metric called  wideband beam gain, which is defined as the minimum  beamforming gain across the entire frequency band of interest, given a target angle of departure (AoD) (or equivalently, angle of arrival (AoA)). Then, the performance of a wideband beamforming codebook for any given target AoD can be evaluated by the maximum wideband beam gain achievable through the best beam. Accordingly, we formulate a new wideband beamforming codebook design problem to jointly optimize a given number of analog beams for improving the worst-case performance of the codebook over all possible AoDs in the spatial domain.

\item To motivate our proposed wideband beamforming design, we first consider the conventional narrowband beamforming codebook as a heuristic solution to our formulated wideband beamforming codebook optimization problem and characterize its performance loss in the wideband THz systems. Then, we solve our formulated new wideband beamforming codebook design problem by dividing  the spatial domain into orthogonal angular zones each associated with one beam, thereby decoupling the original problem into a zone division sub-problem  and a set of beamforming optimization sub-problems each for one zone. For the zone division sub-problem, we propose a bisection method to obtain the optimal boundaries for separating adjacent zones. While for the per-zone-based  beamforming optimization sub-problems, we show that their optimal solutions can be derived by solving a unified beamforming optimization problem.

\item However, the unified beamforming optimization problem  is non-convex and thus difficult to be optimally solved. In light of this, we propose a new approach by combining  the augmented Lagrange method (ALM) and the alternating direction method of multipliers (ADMM). In addition, to accelerate the convergence of the proposed solution, we further propose a piecewise response-vector (PRV) method to initialize all design variables in closed form. Finally, we provide numerical results that show the performance  superiority of our proposed codebook design for wideband analog beamforming compared  to the conventional narrowband beamforming codebook and also  validate our analytical results.
\end{itemize}

The remainder of this paper is organized as follows. Section II presents the system model. Section III introduces  the new performance metric of wideband analog beamforming and formulates the codebook design problem accordingly. Section IV analyzes the performance loss of conventional narrowband beamforming codebook in wideband systems. Section V presents the proposed methods to solve the wideband beamforming  codebook design problem. Section VI evaluates the  performance of the proposed beamforming codebook design via  numerical results. Finally, Section VII concludes the paper.

\section{System Model}
\begin{figure}[t]
\centering
\includegraphics[width=3.5in]{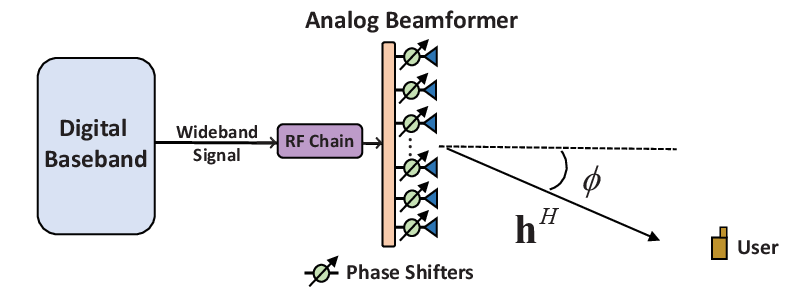}
\caption{A wideband communication system employing analog beamforming in the downlink.}\label{fignarrow}
\vspace{-8pt}
\end{figure}

As depicted in Fig. 1, we consider the downlink transmission of a wideband THz system, where the BS adopts the codebook-based  analog beamforming to transmit a wideband signal to a user\footnote{The proposed analog beamforming method is also applicable to BSs with hybrid beamforming structures for serving multiple users, where each user may be served by an independent RF chain employing the designed codebook for analog beamforming.}. Let ${\cal W}$ and $\phi$ denote the codebook employed by the BS and the AoD from the BS to the user, respectively.  We assume that the user is located in the region specified by $\phi \in [-\pi/2,\pi/2]$.

\subsection{Time-Domain Signal Model}\label{stime}
Let  $s\left( t \right)$ and $B$ represent the BS's baseband signal at time instant $t$ and its bandwidth, respectively. At the BS, the baseband signal $s\left( t \right)$  is first up-converted by an RF chain to a central frequency $f_c$  and processed by an analog beamformer, as shown in Fig. 1. The analog beam is selected from the codebook ${\cal W}$, which is subject to a constant-modulus constraint on each of its elements, i.e.,
\begin{equation}
{\bf{w}}=\frac{1}{\sqrt{N}} {[{e^{ j{\omega _{1}}}},\;{e^{  j{\omega _{2}}}},...,{e^{  j{\omega _{N}}}}]^T}, \;\;{\bf{w}}\in {\cal W},
\end{equation}
where $N$ is the number of BS antennas, $\{\omega _{n}\}_{n=1}^N$ are phase shifts which are assumed to be constant over $B$.  After analog beamforming, the baseband equivalent signal at the $n$-th antenna is 
\begin{equation}\label{x}
x_n(t) = \frac{1}{\sqrt{N}} s(t){e^{j{\omega _n}}}.
\end{equation}
It follows from (\ref{x}) that the phase shifts equivalently tune the phases of the baseband signal. Due to the considerably limited scattering and reflection over the THz band, we assume a line-of-sight (LoS) channel between the BS and the user.  Accordingly, the path loss between the BS and user is given by
\begin{equation}\label{loss}
a(f_c,d_c) = {\frac{c}{{4\pi f_cd_c}}}{e^{-\frac{1}{2}\kappa (f_c)d_c}},
\end{equation}
where $\kappa (f_c)$ and $d_c$ are the medium absorption factor at  frequency $f_c$ and the BS-user distance (assuming a reference point at the BS's antenna array, e.g., at its center), respectively.  Then,  the  baseband channel impulse response at the $n$-th antenna can be expressed as 
\begin{equation}\label{th}
{h_n}(t, \phi) =  a({f_c},{d_c}){e^{ - j2\pi {f_c}{\tau _n}}}\delta (t - {\tau _n}),
\end{equation}
where $\tau _n$ is the transmission delay from the $n$-th antenna to the user, which is a function of $\phi$ (to be specified next). Thus, the received signal at the user in the  time domain can be expressed as 
\begin{equation}\label{timed}
\begin{aligned}
y(t) &= \sum\limits_{n = 1}^N {{h_n}(t, \phi) * {x_n}(t)}  \\
& = \frac{1}{\sqrt{N}} \sum\limits_{n = 1}^N {a({f_c},{d_c}){e^{j({\omega _n} - 2\pi {f_c}{\tau _n})}}s(t - {\tau _n})}.
\end{aligned}
\end{equation}
We assume that the bottom antenna in Fig. 1 is the first antenna and the BS is equipped with a uniform linear array. Then, $\tau _n$ can be expressed as 
\begin{equation}
{\tau _n} = {\tau _1} + (n - 1)\Delta \tau,
\end{equation}
in which ${\tau _1}$ is the time delay from the first antenna to the user, and $\Delta \tau$ is the time delay between two adjacent antennas.   Let $d_a$ denote the spacing between two adjacent antennas. Then, we have
\begin{equation}
\Delta \tau  = \frac{{{d_a}\sin \phi }}{c},  
\end{equation}
where $c$ is the light speed.  For convenience, we assume half-wavelength spacing in this paper, i.e., $d_a=c/(2f_c)$. Based on the above, the antenna array can be equivalently viewed as a multi-path system with different delays over its $N$ paths, and its incurred delay spread is given by
\begin{equation}\label{m8}
T_{\rm{spread}} = |{\tau _N} - {\tau _1}| = \frac{{(N - 1)|\sin \phi| }}{2f_c}.
\end{equation}
It is noted that if $\phi=0$,  there will be no time delay spread. While if $\phi= \pm \pi/2$, the maximum time delay spread will be achieved with $T_{\rm{spread}}  = {(N - 1)\big/2f_c}$.  
Then, if $N$ is sufficiently large (e.g., in the order of tens or even hundreds in THz systems), $T_{\rm{spread}}$ may not be ignored for the resulted inter-symbol interference (ISI) even in our considered  LoS channel setting. To deal with this issue, we can adopt orthogonal frequency division multiplexing (OFDM) modulation and set its cyclic prefix (CP) duration (denoted as ${T_{{\rm{cp}}}}$) larger than the maximum time spread to completely remove ISI, i.e.,
\begin{equation}
{T_{{\rm{cp}}}} >  \frac{{(N - 1)}}{{2{f_c}}}.
\end{equation}


%

\subsection{Frequency-Domain Signal Model}
Next, we characterize the considered system from the frequency domain. By applying Fourier transform to (\ref{th}), the  channel frequency response at the $n$-th antenna can be obtained as 
\begin{equation}
{H_n}(f,\phi) = a({f_c},{d_c}){e^{ - j2\pi {f_c}\tau_n}}{e^{ - j2\pi f\tau_n}}.
\end{equation}
Consequently, the channel frequency response vector at the BS can be written as 
\begin{equation}\label{wdh}
\begin{aligned}
\widetilde {\bf{h}}(f,\phi ) &= {[{H_1}(f,\phi ),{H_2}(f,\phi ),...,{H_N}(f,\phi )]^H}\\
& = a({f_c},{d_c}){e^{j2\pi (f + {f_c}){\tau _1}}} \times \\
& \quad {\left[ {1,{e^{j\pi (1 + \frac{f}{{{f_c}}})\sin \phi }},...,{e^{j(N - 1)\pi (1 + \frac{f}{{{f_c}}})\sin \phi }}} \right]^T}.
\end{aligned}
\end{equation}
As such, the user's baseband received signal in the frequency domain can be expressed as 
\begin{align}\label{fred}
Y(f) &=\widetilde  {\bf{h}}{(f,\phi)^H}{\bf{w}}S(f) = \frac{1}{{\sqrt N }}a({f_c},{d_c}){e^{ - j2\pi (f + {f_c}){\tau _1}}} \times \notag \\
&\qquad \sum\limits_{n = 1}^N {{e^{j[{\omega _n} - (n - 1)\pi (1 + \frac{f}{{{f_c}}})\sin \phi ]}}} S(f),
\end{align}
where $S(f)$ represents the signal component at frequency $f$, $f\in[-\frac{B}{2},	\frac{B}{2}]$. 
Note that for narrowband systems (i.e., $B/2 \ll f_c$), we have $| \frac{f}{f_c}| \approx 0$. Thus, the optimal ${\bf{w}}$ for conventional narrowband systems that maximizes each $|Y(f)|^2$ can be aligned with $\widetilde  {\bf{h}}(0,\phi)$ and is given by
\begin{equation}\label{nar}
{\bf{w}}^{\rm{nar}}( \phi)  = \frac{1}{\sqrt{N}}{[1,{e^{  j\pi\sin \phi  }},...,{e^{  j(N - 1)\pi\sin \phi }}]^T}.
\end{equation}
However, in our considered wideband systems, $B$ is generally not  negligible as compared to $f_c$. As a result, using the narrowband beamforming vector given in  (\ref{nar}) can incur significant performance loss.  This is because 
with a large $B$, the change of $|Y(f)|^2$ over different frequencies cannot be ignored, and the optimal beamforming should cater to  all frequencies over $B$, not only to the central frequency as in the narrowband system.
\begin{remark}
The performance loss by using (\ref{nar})  in the wideband system is due to  the phenomenon of beam squint\cite{bsquint}, where the beam at $f=0$ is steered toward the direction of $\phi $ whereas the beams at other frequencies point to unintended directions rather than $\phi$.
\end{remark}

\section{Wideband Beamforming  Metric and Codebook Design}
In this section, we first define the \emph{wideband beam gain} as a metric to characterize the performance of any analog beamforming in the wideband system for any target AoD.  Then, we formulate an optimization problem  to design the wideband beamforming codebook ${\cal W}$, accounting for all frequencies and all AoDs, i.e., $f \in [-B/2,B/2]$ and $\phi \in [-\pi/2,\pi/2]$.
\subsection{Wideband Beam Gain}
Based on (\ref{fred}), we define the following beam gain at frequency $f$, i.e.,
\begin{equation}\label{ggain}
\begin{aligned}
g(f,\phi,{\bf{w}})&=|{\bf{h}}{(f,\phi )^H}{\bf{w}}|^2\\
&=\left|\frac{1}{\sqrt{N}} \sum\limits_{n = 1}^N {{e^{j[{\omega _n} - (n - 1)\pi(1 + \frac{f}{f_c})\sin \phi ]}}}\right|^2,
\end{aligned}
\end{equation}
where ${\bf{h}}(f,\phi )$ is obtained by normalizing  $ \widetilde {\bf{h}}(f,\phi )$ in (\ref{wdh}) with the path loss, i.e.,
\begin{equation}
{\bf{h}}(f,\phi ) =  {\left[ {1,{e^{  j\pi (1 + \frac{f}{{{f_c}}})\sin \phi }},...,{e^{  j(N - 1)\pi (1 + \frac{f}{{{f_c}}})\sin \phi }}} \right]^T}.
\end{equation}
Next, to describe the performance of any analog beam ${\bf{w}}$ for a given $\phi$, we define the wideband beam gain as
\begin{equation}\label{wg}
G(\phi ,{\bf{w}}) = \mathop {\min }\limits_{f \in [ - \frac{B}{2},\frac{B}{2}]} g(f,\phi ,{\bf{w}}),
\end{equation}
which is the minimum beam gain over the entire frequency band $f\in[-\frac{B}{2},	\frac{B}{2}]$.
\begin{remark}
It is worth mentioning that (\ref{wg}) is relevant to the performance of practical systems. For example, in 5G new radio (NR) system, different frequency subcarriers to one user apply a single modulation and coding scheme (MCS), and thus its performance depends on the minimum effective BS-user channel gain over all subcarriers, i.e., the wideband beam gain defined in  (\ref{wg}). Although the adaptive OFDM scheme can support multiple modulation modes for each subcarrier\cite{ofdm}, it requires a high-cost control link to transmit the modulation allocation information, which rarely appears in practical systems.
\end{remark}

For any target AoD $\phi$, we should search over all the beams in the codebook ${\cal W}$ and select the one that yields the maximum wideband beam gain. Let $L$ and ${\bf{w}}_l$ denote the number of beams and the $l$-th beam in the codebook ${\cal W}$, respectively. Then, we define the \emph{codebook beamforming performance} of ${\cal W}$ for the target AoD $\phi$  as
\begin{equation}\label{cpc}
\Gamma (\phi ,{\cal W}) = \;\mathop {\max }\limits_{l \in \{ 1,2,...,L\} } \;G(\phi ,{{\bf{w}}_l}),
\end{equation}
which indicates the maximum wideband beam gain over $\phi$ achieved by the vectors in ${\cal W}$.

\subsection{Problem Formulation}
Since the AoD of the user is randomly distributed in $[-\pi/2,\pi/2]$, we consider the worst-case beamforming performance of the codebook $ {\cal W}$ over this angle region, i.e.,
\begin{equation}\label{wL}
\begin{aligned}
{\Gamma _{{\rm{worst}}}}(\cal W) &= \mathop {\min }\limits_{\phi  \in [ - \frac{\pi }{2},\frac{\pi }{2}]} \Gamma (\phi,{\cal W} )\\
 &= \mathop {\min }\limits_{\phi  \in [ - \frac{\pi }{2},\frac{\pi }{2}]} \;\mathop {\max }\limits_{l \in \{ 1,2,...,L\} } \;G(\phi ,{{\bf{w}}_l}),\;\;\\
 &= \mathop {\min }\limits_{\phi  \in [ - \frac{\pi }{2},\frac{\pi }{2}]} \;\mathop {\max }\limits_{l \in \{ 1,2,...,L\} } \;\mathop {\min }\limits_{f \in [ - \frac{B}{2},\frac{B}{2}]} {\left| {{\bf{h}}{{(f,\phi )}^H}{{\bf{w}}_l}} \right|^2}.
\end{aligned}
\end{equation}
Accordingly, for the codebook design, we aim to optimize ${\cal W}$ (or its $L$ beams) to maximize ${\Gamma _{{\rm{worst}}}}$, i.e.,
 \[\begin{aligned}
({\rm{P1}}):\;\;&\mathop {\max }\limits_{\{ {{\bf{w}}_l}\} _{l = 1}^L} \; \mathop {\min }\limits_{\phi  \in [ - \frac{\pi }{2},\frac{\pi }{2}]} \;\mathop {\max }\limits_{l \in \{ 1,2,...,L\} }\;\mathop {\min }\limits_{f \in [ - \frac{B}{2},\frac{B}{2}]} |{\bf{h}}{(f,\phi )^H}{\bf{w}}_l|^2\\
&\quad {\rm{s}}.{\rm{t}}.\; |{{\bf{w}}_l}(i)| = \frac{1}{\sqrt{N}},\;l = 1,2,...,L,\;i = 1,2,...,N.
\end{aligned}
\]
It should be mentioned that although the codebook design has been investigated in prior works for narrowband systems (see e.g., the bottom-stage vectors in \cite{hb2}), our considered problem needs to further cater to a continuous wideband in the frequency domain, in addition to the spatial range of $\phi$. Furthermore, the non-convex modulus constraints on the elements of $\{ {{\bf{w}}_l}\} _{l = 1}^L$ make $(\rm{P}1)$ even more difficult to be solved. 

\begin{remark}
The proposed wideband beamforming codebook design can be jointly applied with beam management strategies for serving mobile users, e.g., by employing the beam training and tracking procedure proposed in \cite{3D}.
\end{remark}

\section{Performance Analysis of  Narrowband Beamforming in Wideband Systems}
In this section, we analyze the performance loss of the conventional narrowband beamforming codebook in wideband systems, which can be regarded as a heuristic solution to $(\rm{P}1)$ and also motivates our proposed wideband beamforming codebook design (see Section \ref{wco}). In particular, we unveil two opposite  effects on $\Gamma_{\rm{worst}}$ incurred by changing the number of BS antennas $N$ for the narrowband beamforming codebook, and derive the optimal $N$ that maximizes $\Gamma_{\rm{worst}}$. 
\subsection{Narrowband Beamforming Codebook}
 \begin{figure*}[t]
\centering
\includegraphics[width=6in]{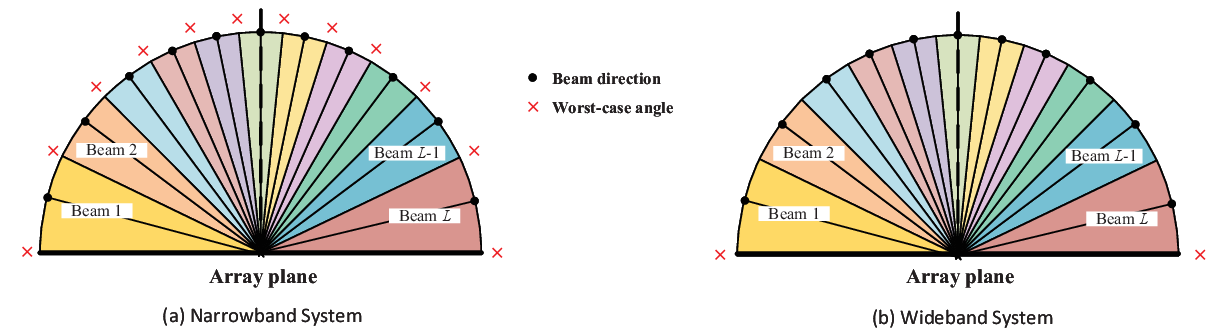}
\caption{Illustration of the angular zones and worst-case user’s AoDs for the narrowband beamforming codebook.}\label{wor}
\vspace{0pt}
\includegraphics[width=7in]{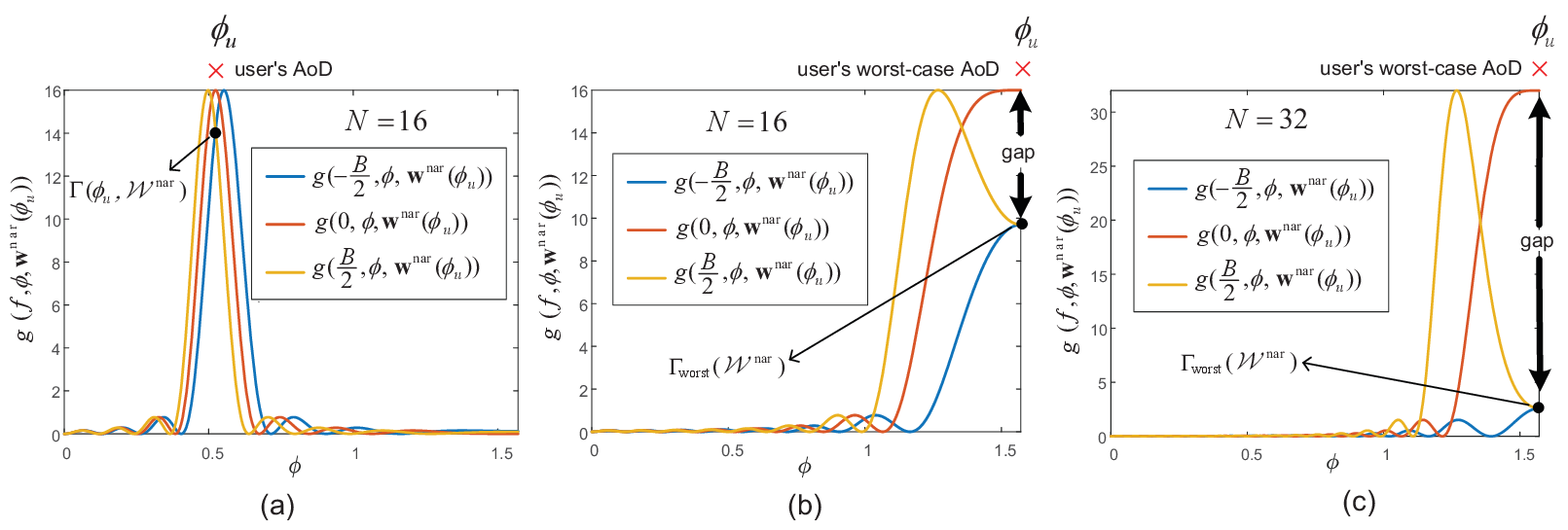}
\caption{Codebook performance by ${\cal W}^{\rm{nar}}$ at (a) a given AoD with $N=16$, (b) the worst-case AoD with $N=16$, and (c) the worst-case AoD with $N=32$.}\label{sam}
\vspace{-8pt}

\end{figure*}

For narrowband systems, the worst-case performance of any analog beamforming codebook  ${\cal W}$ is given by
\begin{equation}\label{np}
\hat\Gamma _{{\rm{worst}}} ({\cal W})= \mathop {\min }\limits_{\phi  \in [ - \frac{\pi }{2},\frac{\pi }{2}]} \mathop {\max }\limits_{l \in \{ 1,2,...,L\} } g(0,\phi ,{{\bf{w}}_l}),
\end{equation}
which is equivalent to setting $B=0 $ in (\ref{wL}). For $L\geq N$ in practice, it can be shown that the optimal beams that maximize the worst-case performance in (\ref{np}) are  given by the following array response vectors\cite{hb2}:
\begin{equation}\label{codebook}
\begin{aligned}
{\bf{w}}_l^{\rm{nar}}  = {\bf{w}}^{\rm{nar}}( \phi_l),\;{\phi _l} = \arcsin \left[ {\frac{{2l - 1}}{L} - 1} \right],\;l = 1,2,...,L.
\end{aligned}
\end{equation}
In particular, the $l$-th beam covers a specific angular zone given by
\begin{equation}\label{zone}
{\Pi _l^{\rm{nar}}} = \left[ {\arcsin \left( { - 1 + \frac{{2(l - 1)}}{L}} \right),\arcsin \left( { - 1 + \frac{{2l}}{L}} \right)} \right].
\end{equation}
More specifically, the zones in narrowband systems are uniformly divided in the sine-AoD space from $-1$ to $1$, and the beam patterns $\{g(0,\phi ,{{\bf{w}}_l})\}_{l=1}^L$ are identical in the $L$ zones, which thus help improve the worst-case performance in (\ref{np}). In particular, as shown in Fig. \ref{wor}(a), the worst-case performance, marked by red cross, occurs when the AoD is at the edge of two adjacent zones, which gives rise to $L+1$ worst-case AoDs with an identical minimum beamforming gain. Given the number of BS antennas $N$ and the number of beams $L$, the worst-case performance by the narrowband beamforming codebook ${\cal W}^{\rm{nar}} \triangleq \{{\bf{w}}_l^{\rm{nar}} \}_{l=1}^L$ in   narrowband systems is given by\cite{hb2}
\begin{equation}
\hat\Gamma _{{\rm{worst}}}({\cal W}^{\rm{nar}}) = \left[\frac{{\sin \left( {N\pi /2L} \right)}}{{\sqrt{N}\sin \left( {\pi /2L} \right)}}\right]^2.
\end{equation}
\subsection{Worst-case Performance by Narrowband  Codebook in Wideband System}
In wideband systems, the conventional narrowband beamforming codebook is far from optimal as the beam patterns, i.e., $\{G(\phi ,{\bf{w}}_l^{\rm{nar}})\}_{l=1}^L$, are different over angular zones. In this subsection, we characterize its worst-case performance in the wideband system case, i.e., ${\Gamma _{{\rm{worst}}}} ({\cal W}^{\rm{nar}})$, for  comparison with our proposed wideband beamforming codebook design in Section \ref{wco}.

\begin{proposition}\label{pp1}
Given the number of BS antennas $N$, the signal  bandwidth $B$, and the number of beams $L$, the worst-case performance of ${\cal W}^{\rm{nar}}$ in wideband system is given by
\begin{align}\label{gm}
&{\Gamma _{{\rm{worst}}}}({\cal W}^{\rm{nar}}) =\\
& \left\{ {\begin{aligned}
&{{{\left[ {\frac{{\sin \left[ {N\pi \left( {2{f_c} + BL} \right)/4{f_c}L} \right]}}{{\sqrt N \sin \left[ {\pi \left( {2{f_c} + BL} \right)/4{f_c}L} \right]}}} \right]}^2},\;{\rm{if}}\; N < \frac{{4{f_c}L}}{{2{f_c} + BL}} }\\
&0,\qquad \qquad \qquad \qquad \qquad \qquad \qquad {\rm{otherwise}},
\end{aligned}} \right. \notag
\end{align}
where the worst-case performance occurs when the user's AoD is $\phi=\pm \pi/2$ (see Fig. {\ref{wor}}(b) marked by red cross).  Based on (\ref{gm}), a necessary condition for the narrowband beamforming codebook ${\cal W}^{\rm{nar}}$  to achieve a non-zero wideband beam gain is given by  
\begin{equation}
\frac{B}{{{f_c}}}  < \frac{4}{N}.
\end{equation}
\end{proposition}
\begin{IEEEproof}
See Appendix A.
\end{IEEEproof}

As shown  in the proof of Proposition 1, ${\Gamma _{{\rm{worst}}}}({\cal W}^{\rm{nar}}) $ monotonically decreases with $B$ and increase with $L$. As such, to improve the worst-case performance, we can decrease $B$ or increase $L$. This is also expected, as decreasing $B$ helps reduce the wideband \emph{beam squint} effect while increasing $L$ helps increase the spatial resolution of the codebook. 

However, it is  difficult to describe the relationship between ${\Gamma _{{\rm{worst}}}}({\cal W}^{\rm{nar}}) $ and $N$. For ease of exposition, we assume that $L$ is infinitely large, such that there exists  a beam ${\bf{w}}^{\rm{nar}}( \phi_u)$ which can perfectly align with the user's AoD, $\phi_u$. In this case, the codebook beamforming performance at $\phi_u$, i.e., $\Gamma (\phi _u , {\cal W}^{\rm{nar}})$ in (\ref{cpc}), is equal to $g(f,\phi _u ,{\bf{w}}^{\rm{nar}}( \phi_u))$ at $f=\pm \frac{B}{2}$ as shown in Fig. {\ref{sam}}(a).  Then, the worst-case performance ${\Gamma _{{\rm{worst}}}}({\cal W}^{\rm{nar}}) $ can be achieved by setting $\phi_u$ in $\Gamma (\phi _u , {\cal W}^{\rm{nar}})$ as the user's worst-case AoDs, i.e.,  $\phi _u=\frac{\pi}{2}$ based on Proposition 1, as shown in Fig. {\ref{sam}}(b). Based on the above, we have
\begin{equation}
{\Gamma _{{\rm{worst}}}}({\cal W}^{\rm{nar}}) =\Gamma \big(\frac{\pi}{2}, {\cal W}^{\rm{nar}}\big)=g\big(\pm \frac{B}{2}, \frac{\pi}{2} ,{\bf{w}}^{\rm{nar}}(  \frac{\pi}{2})\big).
\end{equation}
In this regard, increasing $N$ can improve the beam gain at $f=0$, i.e., $g(0,\frac{\pi}{2},{\bf{w}}^{\rm{nar}}( \frac{\pi}{2}))$, which is beneficial to enhance ${\Gamma _{{\rm{worst}}}}({\cal W}^{\rm{nar}})$. On the other hand,  however, as $N$ increases, the width of each beam in ${\cal W}^{\rm{nar}}$ becomes narrower, and hence its beam gain decreases faster from  $g(0,\frac{\pi}{2} ,{\bf{w}}^{\rm{nar}}( \frac{\pi}{2}))$ to $g(\pm \frac{B}{2},\frac{\pi}{2} ,{\bf{w}}^{\rm{nar}}( \frac{\pi}{2}))$,  which may decrease ${\Gamma _{{\rm{worst}}}}({\cal W}^{\rm{nar}})$. For example, as illustrated in Figs. {\ref{sam}}(b) and {\ref{sam}}(c),  with increasing  $N$ from $16$ to $32$, $g(0,\frac{\pi}{2},{\bf{w}}^{\rm{nar}}( \frac{\pi}{2}))$ increases from $16$ to $32$ accordingly, whereas  the gap between $g(0,\frac{\pi}{2} ,{\bf{w}}^{\rm{nar}}( \frac{\pi}{2}))$ and $g(\pm \frac{B}{2},\frac{\pi}{2} ,{\bf{w}}^{\rm{nar}}( \frac{\pi}{2}))$ becomes larger. Based on the above, there exists two opposite effects on ${\Gamma _{{\rm{worst}}}}({\cal W}^{\rm{nar}}) $ by increasing $N$; thus, there should exist an $N$ which optimally balances the two effects and maximizes ${\Gamma _{{\rm{worst}}}}({\cal W}^{\rm{nar}}) $, as presented in the following proposition.

\begin{proposition}\label{pp2}
The optimal number of BS antennas  that maximizes ${\Gamma _{{\rm{worst}}}}({\cal W}^{\rm{nar}}) $ is given by
\begin{equation}\label{optinn}
\begin{aligned}
{N^*} = & \arg \mathop {\max }\limits_N \;{\Gamma _{{\rm{worst}}}}({\cal W}^{\rm{nar}})\\
&\;\;{\rm{s}}{\rm{.t}}{\rm{.}}\;N \in \left\{ {\left\lfloor {\frac{{1.485{f_c}L}}{{2{f_c} + BL}}} \right\rfloor ,\left\lceil {\frac{{1.485{f_c}L}}{{2{f_c} + BL}}} \right\rceil } \right\},
\end{aligned}
\end{equation}
where the operators $\left\lfloor  \cdot  \right\rfloor $ and $ \left\lceil  \cdot  \right\rceil$ round up and round down its argument to an integer, respectively.
\end{proposition}
\begin{IEEEproof}
See Appendix C. 
\end{IEEEproof}

\section{Wideband Beamforming Codebook Optimization}\label{wco}
As shown in the previous section, the narrowband beamforming codebook may result in significant performance loss in the wideband system. To deal with this issue, we propose a new wideband beamforming codebook design in this section by solving problem $(\rm{P}1)$. Similar to the narrowband beamforming codebook, we also assume $L \geq N$ wideband  beams to respectively cover $L$ consecutive angular zones in the range $\phi \in [-\pi/2,\pi/2]$,  but refine the division of the zones and the beams to mitigate the wideband beam squint effect. Let ${\Pi _l} = \left[ {\phi _{l-1} ,\phi _l} \right]$ denote the \emph{angular zone} covered by the $l$-th beam in the proposed wideband beamforming codebook, with  $\phi _0  =  - \pi /2$ and $\phi _L  =  \pi /2$. Since the $l$-th zone is intended to be covered by ${{\bf{w}}_l}$, we have 
\begin{equation}
\mathop {\max }\limits_{q \in \{ 1,2,...,L\} } \;G(\phi,{{\bf{w}}_q}) = G(\phi ,{{\bf{w}}_l}),\; \forall\;\phi \in \Pi_l.
\end{equation}
Then, the worst-case performance of the codebook in the $l$-th zone can be expressed as
\begin{equation}\label{locw}
\Gamma _{{\rm{worst}}}^l = \mathop {\min }\limits_{\phi  \in {\Pi _l}}\mathop {\max }\limits_{q \in \{ 1,2,...,L\} } \;G(\phi,{{\bf{w}}_q}) =\mathop {\min }\limits_{\phi  \in {\Pi _l}} \;G(\phi ,{{\bf{w}}_l}),
\end{equation}
which we refer to as \emph{local worst-case performance} in the $l$-th zone. Thus, $(\rm{P}1)$ can be decoupled into  $L$ sub-problems, where the $l$-th sub-problem aims to optimize ${{\bf{w}}_l}$ to maximize the local worst-case performance $\Gamma _{{\rm{worst}}}^l $, i.e.,
\begin{equation}
\begin{split}
({\rm{P1}} {\text{-}} l):\;&\mathop {\max }\limits_{{{\bf{w}}_l}} \;\mathop {\min }\limits_{\phi  \in {\Pi _l} } \mathop {\min }\limits_{f \in [ - \frac{B}{2},\frac{B}{2}]} \left| {{\bf{h}}{{(f,\phi )}^H}{{\bf{w}}_l}} \right|^2\\
&\quad {\rm{s}}.{\rm{t}}.\;|{{\bf{w}}_l}(i)| = \frac{1}{\sqrt{N}},\;i = 1,2,...,N.
\end{split}
\end{equation}
As discussed in Section IV, due to the wideband effect,  the uniform division of the zones in the sine space (as in the narrowband beamforming codebook) may not be optimal.  In the following, we first determine the optimal division of the zones for the wideband system, i.e., the optimal $L-1$ boundary points $\{\phi_l\}_{l=1}^{L-1}$  of the $L$ zones. 
Then, under the optimal zone division, we propose a new algorithm to solve $({\rm{P1}} {\text{-}} l)$, $l=1,2, ... ,L$.

\subsection{Optimal Zone Division}\label{zd}
First, it is obvious that for any codebook ${\cal W}$, the local worst-case performance $\Gamma _{\rm{worst}}^l$ in (\ref{locw}) must be no smaller than the global worst-case performance $\Gamma _{\rm{worst}} $ over the entire spatial space, i.e.,
\begin{equation}
\Gamma _{\rm{worst}} \leq \Gamma _{\rm{worst}}^l,\;\;\forall l.
\end{equation}
As such, to enhance $\Gamma _{\rm{worst}}$, it is desired that the local worst-case performance can be identical over all zones, i.e., $\Gamma _{\rm{worst}} = \Gamma _{\rm{worst}}^l$ for all $l$. Next, we show that this is possible by considering an equivalent zone division in the frequency-spatial composite 
domain. Specifically, for $({\rm{P1}}{\text{-}}l)$, we define a frequency-spatial  composite variable
\begin{equation}
\hat f = \frac{{f + {f_c}}}{{{f_c}}}\sin \phi,  \; f \in [ - \frac{B}{2},\frac{B}{2}],\;\phi  \in [\phi _{l-1} ,\phi _l].
\end{equation}
Then, it follows that $\hat f$ is within a continuous \emph{virtual zone} $[\Omega _l^ -,\Omega _l^ +]$, whose arguments are provided as follows. 
\begin{enumerate}
\item If $\phi _{l-1} ,\phi _l \geq 0$, we have
\begin{equation}\label{vir1}
\Omega _l^ -  = \frac{{{f_c} - \frac{B}{2}}}{{{f_c}}}\sin \phi _{l-1},\;\;\Omega _l^ +  = \frac{{{f_c} + \frac{B}{2}}}{{{f_c}}}\sin \phi _l;
\end{equation}

\item If  $\phi _{l-1} <0 ,\phi _l >0$, we have
\begin{equation}\label{vir2}
\Omega _l^ -  = \frac{{{f_c} + \frac{B}{2}}}{{{f_c}}}\sin \phi _{l-1},\;\;\Omega _l^ +  = \frac{{{f_c} + \frac{B}{2}}}{{{f_c}}}\sin \phi _l ;
\end{equation}

\item If $\phi _{l-1} ,\phi _l \leq 0$, we have
\begin{equation}\label{vir3}
\Omega _l^ -  = \frac{{{f_c} + \frac{B}{2}}}{{{f_c}}}\sin \phi _{l-1} ,\;\;\Omega _l^ +  = \frac{{{f_c} - \frac{B}{2}}}{{{f_c}}}\sin \phi _l.
\end{equation}
\end{enumerate}
Then, $({\rm{P1}}{\text{-}}l)$ is equivalent to the following problem with respect to the frequency-spatial domain, i.e.,
\begin{equation*}
\begin{aligned}
({\rm{P2}}{\text{-}}l):\;\;\Gamma _{\rm{worst}}^l=& \mathop {\max }\limits_{{\bf{w}}_l} \mathop {\min }\limits_{\hat f}   \left| {{\bf{h}}{{(\hat f )}^H}{\bf{w}}_l} \right|^2\\
&{\rm{s}}.{\rm{t}}.\;\;\hat f \in [\Omega_- ^l,\Omega_+ ^l],\\
&\;\quad |{\bf{w}}_l(i)| = \frac{1}{\sqrt{N}},\;i = 1,2,...,N,
\end{aligned}
\end{equation*}
where ${\bf{h}}(\hat f ) = {\left[ {1,{e^{  j\pi\hat f }},...,{e^{ j\pi(N - 1)\hat f  }}} \right]^T}$. To ensure that the local worst-case performance is identical for all zones, i.e., $\Gamma _{\rm{worst}}^l$ is identical for all  $l$, we present the following lemma.
\begin{lemma}
Let  ${\bf{w}}_p^{\rm{opt}}$ denote the optimal solution to $({\rm{P2}}{\text{-}}p)$. If $\Omega_+ ^q-\Omega_- ^q=\Omega_+ ^p-\Omega_- ^p$, the optimal value of $({\rm{P2}}{\text{-}}q)$ is always equal to that of  $({\rm{P2}}{\text{-}}p)$, i.e., $\Gamma _{\rm{worst}}^q=\Gamma _{\rm{worst}}^p$. Besides, the optimal solution to $({\rm{P2}}{\text{-}}q)$ is given by  ${\bf{w}}_q^{\rm{opt}} = {\bf{w}}_p^{\rm{opt}} \odot {\bf{h}}(T)$, where $\odot$ represents the Hadamard (element-wise) product and $T=\Omega_- ^q-\Omega_- ^p=\Omega_+ ^q-\Omega_+ ^p$.
\end{lemma}
\begin{IEEEproof}
See Appendix D. 
\end{IEEEproof}

Lemma 1 indicates that the local worst-case performance in the angular zone $\left[ {\phi _{l-1} ,\phi _l} \right]$ is only determined by the \emph{width} of the virtual zone  $[\Omega_- ^l,\Omega_+ ^l]$.\footnote{While there is no overlap between the angular zones,  the associated virtual zones are partially overlapped.} Thus, it suffices to find the boundary points  $\{\phi_l\}_{l=1}^{L-1}$ that lead to the same width of the virtual zones based on (\ref{vir1})-(\ref{vir3}).  To this end,  we propose a recursive method as detailed below.

Specifically,  let the width of all virtual zones be denoted as  $\Delta \Omega=\Omega_+ ^l-\Omega_+ ^l$.  Starting from $\phi _0  =  - \pi /2$, we can successively calculate $\{\phi _l\}_{l=1}^L$  based on (\ref{vir1})-(\ref{vir3}) in terms of $\Delta\Omega$.  In particular, given $\phi _{l-1}$, we can calculate $\phi _{l}$ as
\begin{equation}\label{rule}
{\phi _l} = \left\{ {\begin{aligned}
&\arcsin \left( {\frac{{\Delta \Omega {f_c} + \left( {{f_c} + \frac{B}{2}} \right)\sin {\phi _{l - 1}}}}{{{f_c} - \frac{B}{2}}}} \right),\\
&{\rm{if}}\;{\phi _{l - 1}} <0, \; \Delta \Omega {f_c} + \left( {{f_c} + \frac{B}{2}} \right)\sin {\phi _{l - 1}} \leq 0\\
&\arcsin \left( {\frac{{\Delta \Omega {f_c} + \left( {{f_c} + \frac{B}{2}} \right)\sin {\phi _{l - 1}}}}{{{f_c} + \frac{B}{2}}}} \right),\\
&{\rm{if}}\; {\phi _{l - 1}} <0, \; \Delta \Omega {f_c} + \left( {{f_c} + \frac{B}{2}} \right)\sin {\phi _{l - 1}} > 0\\
&\arcsin \left( {\frac{{\Delta \Omega {f_c} + \left( {{f_c} - \frac{B}{2}} \right)\sin {\phi _{l - 1}}}}{{{f_c} + \frac{B}{2}}}} \right),\\
&{\rm{if}}\;{\phi _{l - 1}} \ge 0.
\end{aligned}} \right.
\end{equation}
Since $\phi _L$ monotonically increases with $\Delta \Omega$, the bisection method can be utilized to find the value of $\Delta \Omega$ that leads to $\phi _L  =  \pi /2$. Then, all division points are determined accordingly.

Next, we provide the following proposition to derive an upper bound on the  worst-case performance of any beamforming codebook in wideband systems.
\begin{proposition}\label{pp3}
The worst-case performance of any wideband beamforming codebook is upper-bounded by
\begin{equation}\label{grml}
{\Gamma _{{\rm{worst}}}} \leq \frac{2}{\Delta \Omega}.
\end{equation}
\end{proposition}
\begin{IEEEproof}
See Appendix E. 
\end{IEEEproof}
It is worth noting that as $\Delta \Omega$ is merely determined by $B$ and $L$, the upper bound in (\ref{grml}) only depends on $B$ and $L$ and is regardless of the number of BS antenna, $N$. Moreover, the performance upper bound in (\ref{grml}) is tight only for an infinite $N$ and arbitrary $\bf{w}$, while $N$ is practically finite and $\bf{w}$ is subject to a constant-modulus constraint in our considered setup.
\subsection{Unifying the Beamforming Optimization Problems}
After determining the division of the zones, we next solve the $L$ sub-problems $({\rm{P2}} {\text{-}} l)_{l=1}^L$. However, in the case of a large $L$, solving all these $L$ sub-problems may incur a high computational complexity.  Next, we develop a unified problem formulation for them, such that we can obtain all solutions through one-time optimization. Specifically, note that the virtual zones $\{[\Omega_- ^l,\Omega_+ ^l]\}_{l=1}^L$ in these sub-problems have an identical width $\Delta \Omega $. Thus, we can simply unify the constraints on $\hat f$ in these $L$ sub-problems as $\hat f \in \left[ { - \frac{{\Delta \Omega }}{2},\frac{{\Delta \Omega }}{2}} \right]$, and retrieve the solutions to $({\rm{P2}} {\text{-}} l)$, $l=1,2,...,L$, based on Lemma 1.  Such a unified problem formulation can be written as
\begin{equation*}
\begin{aligned}
({\rm{P2}}):\;\;& \mathop {\max }\limits_{\bf{w}} \mathop {\min }\limits_{\hat f}   \left| {{\bf{h}}{{(\hat f )}^H}{\bf{w}}} \right|^2\\
&{\rm{s}}.{\rm{t}}.\;\;\hat f \in \left[ { - \frac{{\Delta \Omega }}{2},\frac{{\Delta \Omega }}{2}} \right],\;|{\bf{w}}(i)| = \frac{1}{\sqrt{N}},\;i = 1,2,...,N.
\end{aligned}
\end{equation*}
Let ${\bf{w}}^{\rm{opt}}$ denote the optimal solution to $({\rm{P2}})$. Based on Lemma 1, the beam for the $l$-th zone (i.e., the optimal solution to $({\rm{P2}} {\text{-}} l)$) can be obtained conveniently as 
\begin{equation}
{{\bf{w}}_l} = {{\bf{w}}^{\rm{opt}}} \odot {\bf{h}}\left( {\frac{{\Omega _l^ -  + \Omega _l^ + }}{2}} \right).
\end{equation}
\begin{remark}
It is worth mentioning that  $({\rm{P2}})$  is a general problem and can be applied to other system setups for different purposes. For example, if we aim to design an analog beamformer to form wide main lobe within $\left[\Omega_ -,\Omega_ + \right]$, we can solve $({\rm{P2}})$ by setting $\hat f \in \left[\Omega_ -,\Omega_ + \right]$. Moreover, if we aim to design a wideband analog beam for the user at the AoD of $\phi$, we can set  $\hat f \in \left[\Omega_ -,\Omega_ + \right]$ with
\begin{equation}
{\Omega_ - } = \left\{ {\begin{aligned}
&{\frac{{{f_c} - \frac{B}{2}}}{2f_c}\sin \phi,\;{\rm{ if }}\;\phi  \geq {\rm{0}}}\\
&{\frac{{{f_c} + \frac{B}{2}}}{2f_c}\sin \phi,\;{\rm{ otherwise}}}
\end{aligned}} \right.
\end{equation}
and
\begin{equation}
{\Omega_ + } = \left\{ {\begin{aligned}
&{\frac{{{f_c} + \frac{B}{2}}}{2f_c}\sin \phi,\;{\rm{ if }}\;\phi  \geq {\rm{0}}}\\
&{\frac{{{f_c} - \frac{B}{2}}}{2f_c}\sin \phi,\;{\rm{ otherwise}}}.
\end{aligned}} \right.
\end{equation}
\end{remark}

\subsection{ALM Algorithm}
In this subsection, we propose an ALM algorithm to solve $({\rm{P2}})$. First, to tackle the continuous $\hat f$ in $({\rm{P2}})$, we discretize it over $\left[ { - \frac{{\Delta \Omega }}{2},\frac{{\Delta \Omega }}{2}} \right]$ as $M$ discrete values, i.e.,
\begin{equation}
{\hat f_m} = - \frac{{\Delta \Omega }}{2}+ \frac{{m - 1}}{{M - 1}}\Delta \Omega , \;\;\forall m.
\end{equation}
Then, we define a matrix ${\bf{S}}\in {\mathbb{C}}^{N \times M}$ with its $M$ columns corresponding to the array vectors at these $M$ discrete points, i.e.,
\begin{equation}
{\bf{S}} = \left[ {{\bf{h}}({{\hat f}_1}),{\bf{h}}({{\hat f}_2}),...,{\bf{h}}({{\hat f}_M})} \right].
\end{equation}
To handle the minimum operation and the modulus in the objective function of $({\rm{P2}})$, we introduce two auxiliary variables $t \in \mathbb{R}$ and ${\bf{r}} \in {\mathbb{C}^M}$.  Then, $({\rm{P2}})$ can be transformed into
\begin{equation*}
\begin{aligned}
({\rm{P3}}):\;\;&\mathop {\max }\limits_{t,{\bf{w}},{\bf{r}}} \;\;t\\
&{\rm{ s}}{\rm{.t}}{\rm{. }}\;\;{\mathop{\rm Re}\nolimits} \left\{ {{{\bf{S}}^H}{\bf{w}} \odot {\bf{r}}} \right\} \ge t \cdot {\bf{1}},\\
&\;\;\;\quad|{\bf{w}}(i)| = \frac{1}{\sqrt{N}},\;\forall i,\\
&\;\;\;\quad|{\bf{r}}(m)| = 1,\;\forall m,
\end{aligned}
\end{equation*}
where ${\bf{1}}$ is a vector whose elements are all equal to $1$. As the square root of the beam gain is upper-bounded by $\sqrt{N}$, i.e., $|{\bf{h}}{({{\hat f}_m})^H}{\bf{w}}| \le \sqrt{N}$, the above problem is equivalent to
\begin{equation*}
\begin{aligned}
({\rm{P4}}):\;\;&\mathop {\min }\limits_{{\bf{w}},{\bf{r}},{\bf{y}},{\bf{x}} }\;\;{\left\| {\bf{y}} \right\|_\infty }\\
&{\rm{ s}}{\rm{.t}}{\rm{.  }}\;\;{\bf{y}} = \sqrt{N} \cdot {\bf{r}} - {{\bf{S}}^H}{\bf{w}},\\
&\qquad {\bf{w}} = {\bf{x}},\\
&\qquad |{\bf{x}}(i)| = \frac{1}{\sqrt{N}},\;\forall i,\\
&\qquad |{\bf{r}}(m)| = 1,\;\forall m,
\end{aligned}
\end{equation*}
where ${\bf{y}} \in {\mathbb{C}^M}$ and ${\bf{x}} \in {\mathbb{C}^N}$ are two auxiliary vector variables. Specifically, ${\bf{y}}$ is used to handle the $\rm{Re}\{\cdot\}$ operator and the infinity norm in the objective function can be used to replace $t$; and  ${\bf{x}}$ is used to facilitate the ADMM. Then, the augmented Lagrangian function of ({\rm{P4}}) can be written as
\begin{equation}
\begin{aligned}
&L({\bf{y}},{\bf{w}},{\bf{x}},{\bf{r}},{\bf{u}},{\bm{\lambda }})\\
 &= {\left\| {\bf{y}} \right\|_\infty } + {\mathop{\rm Re}\nolimits} \left\{ {{{\bf{u}}^H}\left( {{\bf{y}} - \sqrt{N} \cdot {\bf{r}} + {{\bf{S}}^H}{\bf{w}}} \right)} \right\}\\
  &\qquad + \frac{{{\rho _1}}}{2}{\left\| {{\bf{y}} - N \cdot {\bf{r}} + {{\bf{S}}^H}{\bf{w}}} \right\|^2}+ {\mathop{\rm Re}\nolimits} \left\{ {{{\bm{\lambda }}^H}\left( {{\bf{w}} - {\bf{x}}} \right)} \right\}\\
 &\qquad + \frac{{{\rho _2}}}{2}{\left\| {{\bf{w}} - {\bf{x}}} \right\|^2} + \mathbb{I}({\bf{x}}) + \mathbb{I}({\bf{r}}),
\end{aligned}
\end{equation}
where $\mathbb{I}(\cdot)$ represents the indicator function, which returns $0$ if the constraint on its argument is satisfied and  returns $\infty$ otherwise, ${\bf{u}}$ and ${\bm{\lambda }}$ are Lagrange multipliers;  $\rho _1$ and $\rho _2$  are penalty factors. Let ${\bf{\bar u}} = {\bf{u}}/{\rho _1} \in {\mathbb{C}^M}$ and ${\bm{\bar \lambda }} ={\bm{ \lambda}} /{\rho _2} \in {\mathbb{C}^N}$. Then, the augmented Lagrangian function can be rewritten as 
\begin{equation}
\begin{aligned}
&L({\bf{y}},{\bf{w}},{\bf{x}},{\bf{r}},{\bf{\bar u}},{\bm{\bar \lambda }})  \\
&={\left\| {\bf{y}} \right\|_\infty } + \frac{{{\rho _1}}}{2}{\left\| {{\bf{y}} - \sqrt{N} \cdot {\bf{r}} + {{\bf{S}}^H}{\bf{w}} + {\bf{\bar u}}} \right\|^2} \\
&\qquad \qquad  + \frac{{{\rho _2}}}{2}{\left\| {{\bf{w}} - {\bf{x}} + {\bm{\bar \lambda }}} \right\|^2} - \frac{{{\rho _1}}}{2}{\left\| {{\bf{\bar u}}} \right\|^2} \\
&  \qquad \qquad    - \frac{{{\rho _2}}}{2}{\left\| {{\bm{\bar \lambda }}} \right\|^2} +  \mathbb{I}({\bf{x}}) + \mathbb{I}({\bf{r}}).
\end{aligned}
\end{equation}
Based on the dual ascent approach and ADMM, the variables can be optimized iteratively. Supposing that $N_{\rm{ite}}$ is the total number of iterations, in the $(n+1)$-th iteration, we successively update the variables as follows.
\begin{subequations}
\begin{align}
&{{\bf{y}}_{n + 1}} = \mathop {\min }\limits_{\bf{y}} L({\bf{y}},{{\bf{w}}_n},{{\bf{x}}_n},{{{\bf{r}}}_n},{{{\bf{\bar u}}}_n},{{{\bm{\bar \lambda }}}_n}),\label{suba}\\
&{{\bf{w}}_{n + 1}} = \mathop {\min }\limits_{\bf{w}} L({{\bf{y}}_{n + 1}},{\bf{w}},{{\bf{x}}_n},{{{\bf{r}}}_n},{{{\bf{\bar u}}}_n},{{{\bm{\bar \lambda }}}_n}),\label{subw}\\
&{{\bf{x}}_{n + 1}} = \mathop {\min }\limits_{\bf{x}} L({{\bf{y}}_{n + 1}},{{\bf{w}}_{n + 1}},{\bf{x}},{{{\bf{r}}}_n},{{{\bf{\bar u}}}_n},{{{\bm{\bar \lambda }}}_n}),\\
&{{{\bf{r}}}_{n + 1}} = \mathop {\min }\limits_{{\bf{r}}} L({{\bf{y}}_{n + 1}},{{\bf{w}}_{n + 1}},{{\bf{x}}_{n + 1}},{\bf{r}},{{{\bf{\bar u}}}_n},{{{\bm{\bar \lambda }}}_n}),\label{subb}\\
&{{{\bf{\bar u}}}_{n + 1}} = {{{\bf{\bar u}}}_n} + {\beta _1}\left( {{{\bf{y}}_{n + 1}} - N \cdot {{\bf{r}}_{n + 1}} + {{\bf{S}}^H}{{\bf{w}}_{n + 1}}} \right),\\
&{{{\bm{\bar \lambda }}}_{n + 1}} = {{{\bm{\bar \lambda }}}_n} + {\beta _2}\left( {{{\bf{w}}_{n + 1}} - {{\bf{x}}_{n + 1}}} \right),
\end{align}
\end{subequations}
where ${\beta _1}$ and ${\beta _2}$ are the step sizes of the dual ascent. The solutions to (\ref{suba})-(\ref{subb}) are presented in the following.
\subsubsection{Update ${\bf{y}}$}
By discarding the irrelevant terms,  it is equivalent to solving 
\begin{equation}\label{m43}
\mathop {\min }\limits_{\bf{y}} \;\;{\left\| {\bf{y}} \right\|_\infty } + \frac{{{\rho _1}}}{2}{\left\| {{\bf{y}} - {{\bf{c}}_n}} \right\|^2},
\end{equation}
with ${{\bf{c}}_n} = \sqrt{N} \cdot {{\bf{r}}_n} - {{\bf{S}}^H}{{\bf{w}}_n} - {{\bf{\bar u}}_n}$. To solve problem (\ref{m43}), we compute 
\begin{equation}
{\alpha ^*} = \arg \mathop {\min }\limits_{\alpha  \in \mathbb{R}^+ } \;\;\alpha  + \frac{{{\rho _1}}}{2}{\sum\limits_{m = 1}^M {\left( {\alpha  - \left| {{{\bf{c}}_n}(m)} \right|} \right)} ^2},
\end{equation}
which is a quadratic minimization problem and the optimal solution is given by
\begin{equation}
{\alpha ^*} = \max \left\{ {\frac{{{\rho _1}\sum\nolimits_{m = 1}^M {\left| {{{\bf{c}}_n}\left( m \right)} \right| - 1} }}{{M{\rho _1}}},0} \right\}.
\end{equation}
By applying the element-wise truncation of ${\bf{y}}$ to the region of
\begin{equation}
{\bf{y}}(m) \in \left\{ {u \in \left| {{\rm{ }}\left| u \right| \le {\alpha ^*}} \right.} \right\},\forall m,
\end{equation}
we obtain the solution to (\ref{m43}) as
\begin{equation}\label{sy}
{\bf{y}}(m) = \left\{ {\begin{aligned}
&{{{\bf{c}}_n}(m),\;\;\;\qquad {\rm{      if }}\;\;|{c_n}(m)| \le {\alpha ^*}}\\
&{{\alpha ^*}{e^{j\arg \left[ {{{\bf{c}}_n}(m)} \right]}},\;\;{\rm{    otherwise  }}},
\end{aligned}\;\;\forall m.} \right.
\end{equation}
\subsubsection{Update ${\bf{w}}$}
The objective function is convex with respect to ${\bf{w}}$, which is given by
\begin{equation}\label{m48}
\begin{aligned}
&L({{\bf{y}}_{n + 1}},{\bf{w}},{{\bf{x}}_n},{{\bf{r}}_n},{{{\bf{\bar u}}}_n},{{{\bm{\bar \lambda }}}_n})\\
 &= {\left\| {\bf{y}} \right\|_\infty } +\frac{{{\rho _1}}}{2}{\left\| {{{\bf{y}}_{n + 1}} - \sqrt{N} \cdot {{\bf{r}}_n} + {{\bf{S}}^H}{\bf{w}} + {{{\bf{\bar u}}}_n}} \right\|^2}  \\
 &\qquad \qquad  + \frac{{{\rho _2}}}{2}{\left\| {{\bf{w}} - {{\bf{x}}_n} + {{{\bm{\bar \lambda }}}_n}} \right\|^2} - \frac{{{\rho _1}}}{2}{\left\| {{{{\bf{\bar u}}}_n}} \right\|^2}\\
 &\qquad \qquad  - \frac{{{\rho _2}}}{2}{\left\| {{{{\bm{\bar \lambda }}}_n}} \right\|^2} + \mathbb{I}({{\bf{x}}_n}) + \mathbb{I}({{\bf{r}}_n}).
\end{aligned}
\end{equation}
The derivative of (\ref{m48}) with respect to ${{\bf{w}}^*}$ is given by
\begin{equation}
\begin{aligned}
\frac{{\partial L}}{{\partial {{\bf{w}}^*}}} =& \frac{{{\rho _1}}}{2}\left[ {{\bf{S}}{{\bf{S}}^H}{{\bf{w}}} - {\bf{S}}\left( {\sqrt{N} \cdot {{\bf{r}}_n} - {{{\bf{\bar u}}}_n} - {{\bf{y}}_{n + 1}}} \right)} \right] \\
& \qquad  \qquad  + \frac{{{\rho _2}}}{2}\left( {{{\bf{w}}} - {{\bf{x}}_n} + {{{\bm{\bar \lambda }}}_n}} \right).
\end{aligned}
\end{equation}
By setting the derivative to zero, it can be shown  that
\begin{equation}\label{w51}
\begin{aligned}
&{{\bf{w}}} = {\left( {{\rho _1}{\bf{S}}{{\bf{S}}^H} + {\rho _2}{\bf{I}}} \right)^{ - 1}} \times \\
&\quad \left[ {\rho _1}{\bf{S}}\left( {\sqrt{N} \cdot {{\bf{r}}_n} - {{{\bf{\bar u}}}_n} - {{\bf{y}}_{n + 1}}} \right) + {\rho _2}\left( {{{\bf{x}}_n} - {{{\bm{\bar \lambda }}}_n}} \right) \right].
\end{aligned}
\end{equation}
\subsubsection{Update ${\bf{x}}$}
It is equivalent to solving
\begin{equation}\label{m51}
\begin{aligned}
&\mathop {\min }\limits_{\bf{x}}  \left\| {{{\bf{w}}_{n + 1}} - {\bf{x}} + {{{\bm{\bar \lambda }}}_n}} \right\|_{}^2\\
&\quad {\rm{  s}}{\rm{.t}}{\rm{.}}\;\;\left| {{\bf{x}}(i)} \right| = \frac{1}{\sqrt{N}},\;\forall i.
\end{aligned}
\end{equation}
It is easy to verify that the optimal solution to (\ref{m51}) is given by
\begin{equation}
{\bf{x}}\left( i \right) = {\frac{1}{\sqrt{N}}e^{j\arg \left[ {{{\bf{w}}_{n + 1}}(i) + {{{\bm{\bar \lambda }}}_n}(i)} \right]}}.
\end{equation}

\subsubsection{Update ${\bf{r}}$}
It is equivalent to solving
\begin{equation}
\begin{aligned}
&\mathop {\min }\limits_{\bf{r}} {\left\| {{{\bf{y}}_{n + 1}} + {{\bf{S}}^H}{{\bf{w}}_{n + 1}} + {{{\bf{\bar u}}}_n} - \sqrt{N} \cdot {\bf{r}}} \right\|^2}\\
&\quad {\rm{  s}}{\rm{.t}}{\rm{.}}\;\;\left| {{\bf{r}}(m)} \right| = 1,\forall m.
\end{aligned}
\end{equation}
Similarly, the optimal solution is given by
\begin{equation}\label{61}
{\bf{r}}\left( m \right) = {e^{j\arg \left[ {{{\bf{y}}_{n + 1}}(m) + {\bf{S}}{{(:,m)}^H}{{\bf{w}}_{n + 1}} + {{{\bf{\bar u}}}_n}(m)} \right]}},
\end{equation}
where ${\bf{S}}{{(:,m)}}$ is the $m$-th column of matrix $\bf{S}$.
\subsection{Initializations, Parameter Setup, and Complexity}
Our proposed ALM algorithm can converge to the local optimal value of $L({\bf{y}},{\bf{w}},{\bf{x}},{\bf{r}})$ by choosing proper penalty factors and step sizes of the dual ascent.  As $({\rm{P4}})$ is a non-convex optimization problem, the global optimality cannot be guaranteed in general and the initialization has a crucial impact on the converged solution. In this paper, we propose a PRV method to provide closed-form initial values for  $\bf{w}$ and $\bf{x}$.  The PRV method is inspired by the codebook design in  \cite{xiaotvt}, which proposed a sub-array combination scheme to construct the wide beams.

\begin{figure}[t]
\centering
\includegraphics[width=2in]{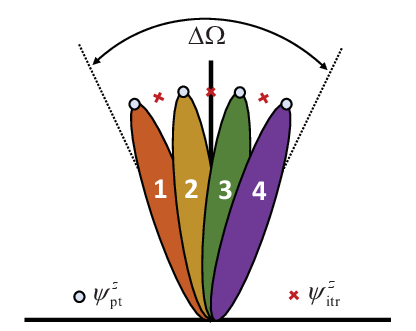}
\caption{Illustration of the sub-array combination scheme with $Z=4$.}\label{z4}
\vspace{-12pt}
\end{figure}

Specifically, we aim to find a beam pointing to direction of $0$ with beamwidth $\Delta \Omega$.  If $\Delta \Omega$ is smaller than $2/N$, we use an array response vector to form the beam. Otherwise, we divide the array into $Z$ identical sub-arrays, i.e., $N=ZN_s$,  each steering a beam (by array response vector with $N_s$ elements) uniformly pointing to all directions within $\left[ { - \Delta \Omega /2,\Delta \Omega /2} \right]$, as shown in Fig. \ref{z4}.  Since the maximum achievable beamwidth  by $Z$ sub-arrays is $2Z/N_s=2Z^2/N$ \cite{xiaotvt},  the minimum $Z$ satisfies
\begin{equation}
\frac{2}{N}{Z^2} \ge \Delta \Omega \;\Rightarrow\; Z \ge \sqrt {\frac{{\Delta \Omega N}}{2}}.
\end{equation}
These $Z$ sub-arrays give rise to $Z$ beam pointing directions and $Z-1$ beam intersection directions, as illustrated in Fig. \ref{z4} via gray circles and red crosses, respectively, and expressed as
\begin{equation}
\begin{split}
\psi _{{\rm{pt}}}^z &=  - \frac{{\Delta \Omega }}{2} + \frac{{(2z - 1)\Delta \Omega }}{{2Z}},\;z = 1,2,...,Z,\\
\psi _{{\rm{itr}}}^z &=  - \frac{{\Delta \Omega }}{2} + \frac{{z\Delta \Omega }}{Z},\;z = 1,2,...,Z - 1.
\end{split}
\end{equation}
 Then, the beam can be constructed as
\begin{equation}
{{\bf{w}}_{\Delta \Omega }} = \frac{1}{\sqrt{N}} \left[ {\begin{aligned}
&{{e^{j{\theta _1}}}{{{\bf{a}}}_{{N_s}}}(\psi _{{\rm{pt}}}^1)}\\
&{{e^{j{\theta _2}}}{{{\bf{a}}}_{{N_s}}}(\psi _{{\rm{pt}}}^2)}\\
&\qquad \;\; \vdots \\
&{{e^{j{\theta _Z}}}{{{\bf{a}}}_{{N_s}}}(\psi _{{\rm{pt}}}^Z)}
\end{aligned}} \right],
\end{equation}
where ${{\bf{a}}_{N_s}}(\psi ) = {\left[ {1,{e^{ - j\pi \psi }},...,{e^{ - j\pi (N_s - 1)\psi }}} \right]^T}$ denotes the array response vector with $N_s$ elements and we  introduce a set of variables $\{{e^{j{\theta _z}}}\}_{z=1}^Z$ to provide additional degrees of freedom (DoF) for adjusting the beam pattern. Note that in the case without these DoF, i.e., $\theta _z=0$ for all $z$, the generated wide beam will have severe trenches in the directions of  beam intersections\cite{arv}. Thus, we aim to optimize the phase shifts $\{\theta _z\}_{z=1}^Z$ to maximize the beam gains in the directions of  $\{\psi _{{\rm{itr}}}^z\}_{z=1}^Z$. The optimal solutions are given by
\begin{equation}\label{thetaclose}
{\theta _z} = \frac{{(Z - z + 1){N_s} - 1}}{{2Z}}(z - 1)\pi \Delta \Omega,
\end{equation}
for which the detailed derivation is given in Appendix F.  As a result, the initializations of $\bf{w}$ and $\bf{x}$ are given by 
\begin{equation}\label{45w}
{\bf{w}} = {\bf{x}} ={{\bf{w}}_{\Delta \Omega }}.
\end{equation}

\renewcommand{\algorithmicrequire}{\textbf{Input:}}
\renewcommand{\algorithmicensure}{\textbf{Output:}}
\begin{algorithm} 
	\caption{Proposed Wideband Beamforming Codebook Design}    
	 \label{xx}       
	\begin{algorithmic}[1] 
	\Require $f_c$, $B$, $N$, $L$, $M$, $N_{\rm{ite}}$.
	    \State Calculate $\Delta \Omega$ and $\{ \Omega _l^-/\Omega _l^+\}_{l}^L$ based on Section \ref{zd}.
	     \State Let ${\bf{S}} = \left[ {{\bf{h}}({{\hat f}_1}),{\bf{h}}({{\hat f}_2}),...,{\bf{h}}({{\hat f}_M})} \right]$, where  $ {\hat f_m} = - \frac{{\Delta \Omega }}{2}+ \frac{{m - 1}}{{M - 1}}\Delta \Omega$ and ${\bf{h}}(\hat f_m ) = {\left[ {1,{e^{  j\pi\hat f_m}},...,{e^{ j\pi(N - 1)\hat f_m }}} \right]^T}$ for all $m$.
    \State Initialize ${\bf{w}}$ and ${\bf{x}}$ based on (\ref{45w}) and set ${\bf{\bar u}}={\bm{\bar \lambda}}={\bf{0}}$; 
     \State   Initialize ${\bf{r}}={{{\bf{S}}^H}{\bf{w}}}$ and normalize its entries as ${\bf{r}}\left( m \right) = \frac{{\bf{r}}\left( m \right)}{\sqrt{N} \left| {{\bf{r}}\left( m \right)} \right|} $ for all $m$.
        \State {\bf{for}}  {$n =1 :  N_{\rm{ite}}$} {\bf{do}} 
        \State \;\; Let ${{\bf{c}}} = N \cdot {{\bf{r}}} - {{\bf{S}}^H}{{\bf{w}}} - {{\bf{\bar u}}}$.
   \State  \;\;   Let ${\alpha ^*} = \max \left\{ {\left( {{\rho _1}\sum\nolimits_{m = 1}^M {\left| {{{\bf{c}}}\left( m \right)} \right| - 1} } \right)\Big/M{\rho _1},0} \right\}.$
     \State \;\; Calculate ${\bf{y}}$ based on (\ref{sy}).
    \State \;\; Calculate ${{\bf{w}}}$ based on (\ref{w51})
    \State \;\; Calculate ${\bf{x}} = {\bf{w}} + {\bm{\bar \lambda }}$ and normalize its entries as ${\bf{x}}(i) = \frac{ {\bf{x}}(i)}{\sqrt{N}\left| {{\bf{x}}(i)} \right|}  ,\;\forall i$.
     \State \;\; Calculate ${\bf{r}} = {\bf{y}} + {{\bf{S}}^H}{\bf{w}} + {\bf{\bar u}}$ and normalize its entries as ${\bf{r}}\left( m \right) = \frac{{\bf{r}}\left( m \right)}{\left| {{\bf{r}}\left( m \right)} \right|},\;\forall m$.
   \State {\bf{end for}} 
      \State Obtain ${{\bf{w}}}^*={\bf{x}}$ as the solution to $({\rm{P2}})$.
   \State {\bf{for}}  {$l =1 : L$} {\bf{do}} 
\State   \;\; Calculate the $l$-th beam as ${{\bf{w}}_l} = {{\bf{w}}^*} \odot {\bf{h}}\left( {\frac{{\Omega _l^ -  + \Omega _l^ + }}{2}} \right).$
   \State {\bf{end for}} 
    \Ensure  The wideband beamforming codebook $\left\{{{\bf{w}}}_1, {{\bf{w}}}_2,...,{{\bf{w}}}_L\right\}$.
   \end{algorithmic} 
\end{algorithm}

Moreover, we initialize ${\bf{\bar u}}$ and ${\bm{\bar \lambda}}$ as  zero vectors. Following (\ref{61}), the entries of ${\bf{r}}$ are initialized as ${\bf{r}}\left( m \right) =  {e^{j\arg \left[ {{\bf{S}}{{(:,m)}^H}{{\bf{w}}}} \right]}}$ for all $m$.  The detailed steps of the wideband beamforming codebook design are summarized in Algorithm 1. In particular, we obtain the optimal zone division in step 1. From steps 2 to 13, we invoke the ALM algorithm to solve the unified problem $({\rm{P2}})$. From steps 14 to 16, we calculate the beams based on Lemma 1.  It can be shown that the overall computational complexity of Algorithm 1 is $\mathcal{O}(N^2M N_{\rm{ite}}+NL)$.  
\begin{figure*}[t]
\centering
\includegraphics[width=6.8in]{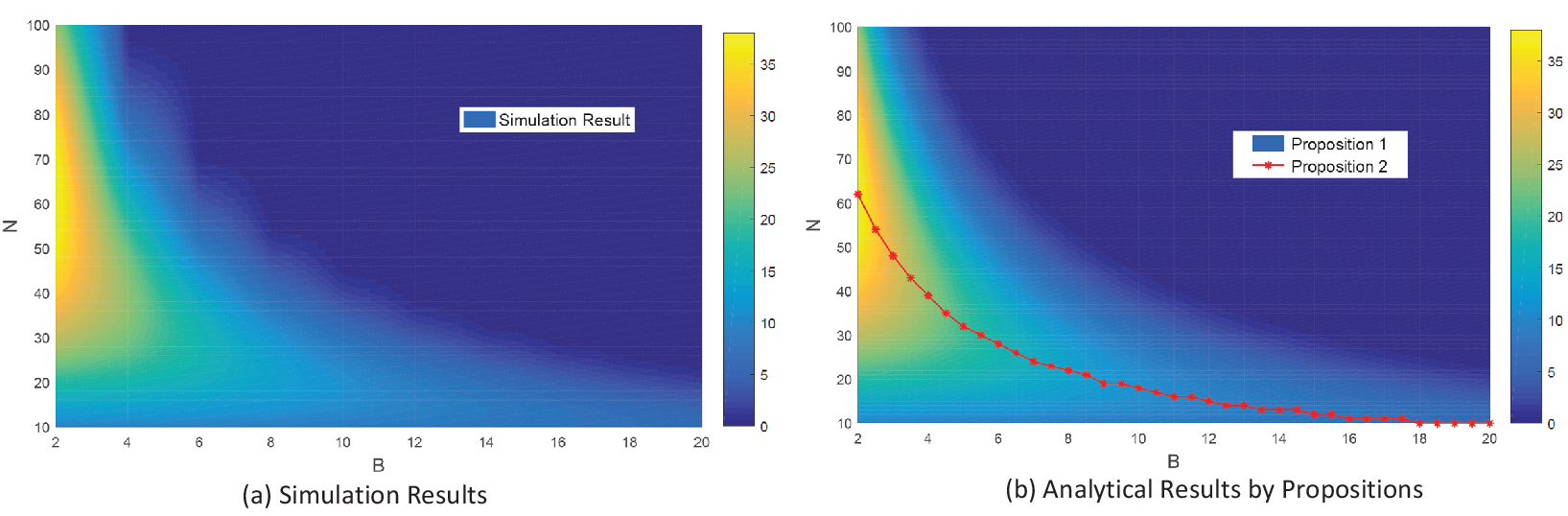}
\caption{The worst-case performance of narrowband beamforming codebook versus $N$ and $B$, where $L=200$ and $f_c=140$ GHz.}\label{co1}
\vspace{4pt}
\includegraphics[width=6.8in]{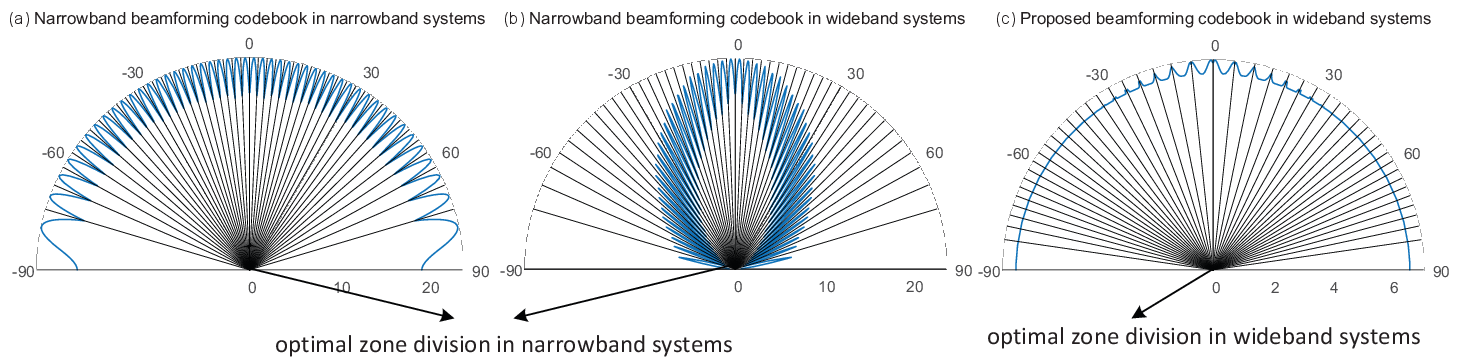}
\caption{Codebook performance and zone division with $N=24$ and $L=48$.}\label{co2}
\vspace{-12pt}
\end{figure*}

\section{Numerical Results}
In this section, we present  numerical results to demonstrate the performance by the conventional narrowband beamforming codebook and our proposed wideband beamforming codebook. In 5G beyond and future THz communication systems, an array-of-subarray hybrid beamforming structure is anticipated to be the prevalent approach in massive MIMO. In this paper, our focus is on the wideband codebook for a single subarray, where a one-dimensional array with dozens of antennas constitutes a logical and practical choice. For instance, if considering a hybrid beamforming architecture with four subarrays, each being a Uniform Planar Array (UPA) with $32$ antennas in one dimension, the total number of antennas is $4 \times 32 \times 32 = 4096$, which is sufficient to compensate for the propagation loss. In ALM, we set $\rho _1=\rho _2=1$ (which control the initial penalty),  $\beta_1=\beta_2=10^{-3}$ (which control the penalty increment), and  $M=2N$.  The total number of iterations $N_{\rm{ite}}$ can be set empirically (we set $50$) or determined by some converge criterion, i.e., the iteration is terminated when ${\left\| {{\bf{y}} - \sqrt{N} \cdot {\bf{r}} + {{\bf{S}}^H}{\bf{w}}} \right\|_2}$  is smaller than a preset threshold $\epsilon$.   For the narrowband beamforming codebook, we show its worst-case performance  via both Monte Carlo simulations and the theoretical results in Proposition 1. In the Monte Carlo simulation, we randomly generate a large number of user AoDs and calculate their corresponding wideband beam gains. Then, the lowest beam gain among them is obtained as the worst-case performance. 

\begin{figure*}[t]
\centering
\includegraphics[width=7in]{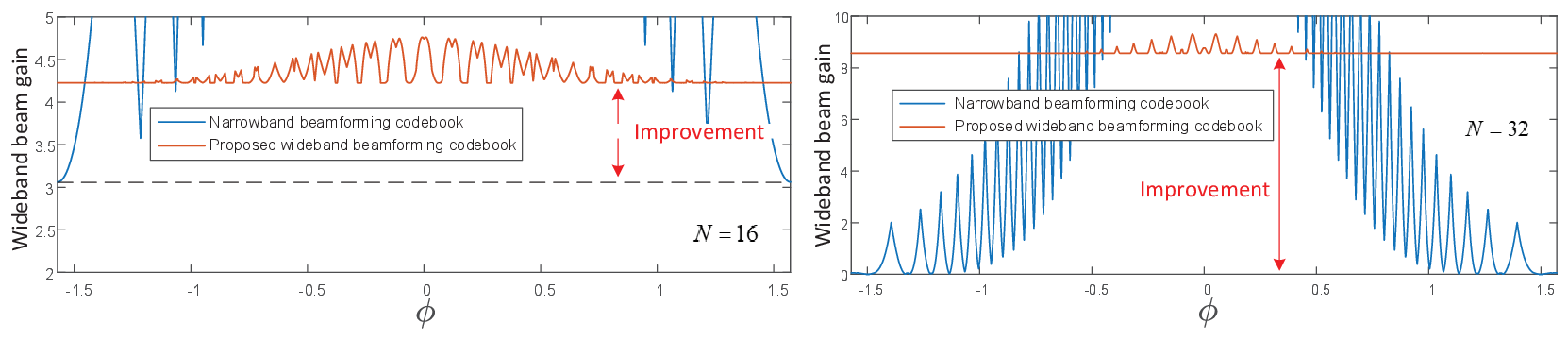}
\caption{The performance improvement of the proposed wideband beamforming codebook with $N=16$ and $N=32$.}\label{co3}
\vspace{12pt}
\includegraphics[width=7in]{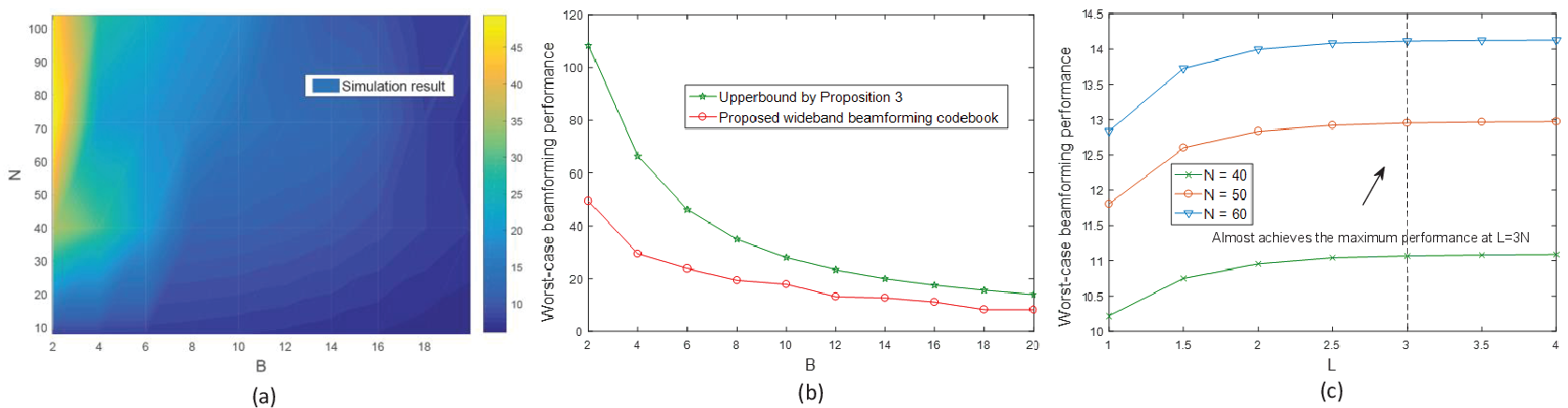}
\caption{The worst-case performance of our proposed wideband beamforming codebook: (a) versus $N$ and $B$, (b) versus $B$ under the optimal $N$, and (c) versus $L$ with $B=10$ GHz.}\label{co4}
\vspace{-12pt}
\end{figure*}First, Fig. \ref{co1} shows the worst-case performance of the  narrowband beamforming codebook versus the number of BS antennas $N$ and the signal bandwidth $B$ (in GHz), with $L = 200$, $f_c = 140$ GHz. In the Monte Carlo simulation, we generate $1000$ user AoDs. It is observed from  Fig. \ref{co1}  that the worst-case performance obtained by Monte Carlo simulations is quite  consistent with that by Proposition \ref{pp1}. It is also observed that the worst-case performance monotonically decreases with $B$, and first increases and then decreases with $N$. This is consistent with our analysis  in Proposition \ref{pp1}. Furthermore, based on Proposition \ref{pp2}, we also plot  the optimal values of $N$ that achieve the best worst-case codebook beamforming performance versus the bandwidth $B$, as marked by asterisk. It is observed that for each given $B$, the highest wideband beam gain incurs at the position of the asterisk, which indicates that the optimum worst-case performance is achieved, thus validating Proposition \ref{pp2}.  


To better visualize the codebook beamforming performance, we plot  the beam patterns of all beamforming vectors in the narrowband and our proposed wideband codebooks in Fig. \ref{co2}, with $f_c = 140$ GHz, $B = 10$ GHz, $N = 24$ and $L = 48$. It is observed that for the narrowband beamforming codebook applied in the narrowband system, as shown in Fig. \ref{co2}(a), its beam pattern becomes increasingly narrower when the AoD changes from the two sides ($\pm90^\circ$) to the middle ($0^\circ$). Accordingly, the optimal zone division also becomes narrower from the two sides to the middle, which ensures that the beam gain in the intersection direction is the same, thereby maximizing the worst-case  performance. However, for the narrowband beamforming codebook applied in the wideband systems, as shown in Fig. \ref{co2}(b), its achieved wideband beam gain is observed to decrease significantly as the AoD changes from $0^\circ$ to $\pm90^\circ$. Particularly, in the zones at the two sides, the wideband beam gain for many AoDs is quite small. This is because the codebook performance in wideband systems is affected by the beam squint effect, which becomes more significant as the AoD changes from $0^\circ$ to $\pm90^\circ$. Finally, Fig. \ref{co2}(c) shows the performance of our proposed wideband beamforming codebook and its corresponding zone division. It is observed that with the proposed wideband beamforming codebook, as the AoD changes from $0^\circ$ to $\pm90^\circ$, the zone width first decreases and then increases, i.e., the optimal zone division is relatively wide in the middle and the two sides, and relatively narrow in the other parts. In addition, our proposed beamforming codebook is observed to achieve the same local worst-case performance in each zone, thereby ensuring the global worst-case performance among all zones.

Fig. \ref{co3} shows the wideband beam gain by our proposed wideband beamforming codebook over the  narrowband beamforming codebook versus the user AoD $\phi$, with $f_c = 140$ GHz, $B = 10$ GHz, $N = 16$ or $32$, and $L = 2N$. For each AoD considered, we show the wideband beam gains achieved by the best beams in the proposed and narrowband beamforming codebooks, respectively. It is observed that when $N = 16$, the wideband beam gain by the narrowband beamforming codebook reaches its minimum value (approximately $3.05$) at $\phi=\pm90^\circ$. In contrast, our proposed beamforming codebook can achieve the minimum value of approximately $4.22$, thus greatly improving over the conventional one. It is also observed that when $\phi$ approaches $0^\circ$, the wideband beam gain becomes much higher than the minimum value of $4.22$. However, when $\phi$ approaches $\pm 90^\circ$, the wideband beam gain is almost identical to $4.22$. This is because due to the beam squint effect, the wideband beam gain $G(\phi ,{\bf{w}})$ in (\ref{wg}) can be interpreted as the minimum beam gain at $f=0$ over a range of AoDs around $\phi$.  As $\phi$ changes from $\pm90^\circ$ to $0^\circ$, the range becomes narrower. Hence, when $\phi$ approaches  $\pm90^\circ$, the wideband beam gain changes very slowly since the minimum beam gain among a wide range of AoDs is the same.  In contrast, when $\phi$  approaches $0^\circ$, the wideband beam gain changes fast and similarly to the narrowband beam gain. When $N = 32$, it is observed that the worst-case wideband beam gain by the narrowband beamforming codebook over all AoDs considered has already dropped to $0$, since the beam squint effect becomes more severe as $N$ increases. In contrast, the worst-case wideband beam gain by the proposed beamforming codebook can increase from $4.22$ to $8.4$, which shows the significant improvement over the narrowband beamforming codebook under a large $N$, which is usually the case for future THz wireless system.

Finally, Fig. \ref{co4}(a) shows the worst-case performance of the proposed wideband beamforming codebook versus $N$ (from $10$ to $140$) and $B$ (from $2$ GHz to $20$ GHz), where $L = 200$ and $f_c = 140$ GHz. It is observed that the worst-case performance of the proposed wideband beamforming codebook decreases with $B$, but first increases and then decreases with $N$, which shows a similar trend to the narrowband beamforming codebook as discussed in Section III. Hence, there should also exist an optimal $N$ to maximize the worst-case performance by the proposed beamforming codebook. As compared to the worst-case performance by the narrowband beamforming codebook over the same $N$-$B$ region, as shown in Fig. \ref{co1}, the proposed wideband beamforming codebook significantly outperforms it over a large part of the region. Furthermore, Fig. \ref{co4}(b)  shows the worst-case performance of the proposed wideband beamforming codebook with the optimal $N$ (obtained via exhaustive search) and the performance upper bound given in Proposition \ref{pp3} versus $B$. It is observed that although increasing $B$ degrades the worst-case performance, the rate of degradation becomes slower, which presents a similar trend as the curve in Fig. \ref{co1} as $B$ increases.  Also, the gap to the performance upper bound decreases with $B$.  Fig. \ref{co4}(c)  shows the codebook beamforming performance versus the number of beams $L$ with $B=10$ GHz. It is observed that the codebook beamforming performance improves with $L$, as expected, but the growth rate becomes slower with increasing $L$. In particular, when $L \ge 3N$, further increasing $L$ can barely improve the codebook beamforming performance, which implies that a sufficiently high resolution has been reached. Note that a larger $L$ results in a higher training overhead for beam training in practical THz communication systems. Thus, the number of beams should be properly selected to achieve the best trade-off between the beamforming gain and the training overhead.

\section{Conclusion}
In this paper, we considered a wideband THz systems adopting analog beamforming. By introducing a new metric of wideband beam gain, we analyzed the performance loss of the conventional narrowband beamforming codebook and drew useful insights. In addition, we formulated a wideband beamforming codebook design problem to maximize the worst-case performance of the codebook accounting for both frequency and spatial domains. To solve such a challenging problem, we decomposed it into a zone division sub-problem and a set of beamforming optimization sub-problems and proposed efficient solutions to them. Numerical results demonstrated that our proposed zone division design can adapt to the variation of beam width and mitigate the beam squint effect effectively, thereby yielding significant performance gains over the conventional narrowband beamforming codebook.  This paper can be extended to several interesting directions for future work. For example, the proposed metric of wideband beam gain can also be applied to other scenarios such as near-field THz channels, THz integrated sensing and communication (ISAC), and THz passive beamforming (e.g., with intelligent reflecting surface (IRS)), etc.
\begin{appendices}    

\section{Proof of Proposition 1}
Assume that ${\bf{w}} = {\bf{a}}(\phi _m)$ is the best beam selected from the narrowband beamforming codebook given in (\ref{codebook}). Then, the wideband beam gain can be written in closed form as 
\begin{align}\label{close}
&G(\phi ,{\bf{a}}({\phi _m})) = \\
&\left| {\frac{{\sin \frac{{N\pi }}{{2{f_c}}}\left[ {{f_c}|\sin {\phi _m} - \sin \phi | + \frac{B}{2}|\sin \phi |} \right]}}{{\sqrt{N}\sin \frac{\pi }{{2{f_c}}}\left[ {{f_c}|\sin {\phi _m} - \sin \phi | + \frac{B}{2}|\sin \phi |} \right]}}} \right|^2. \notag
\end{align}
The details on how to derive (\ref{61}) are deferred to Appendix B. One can prove that the wideband beam gain decreases from $N$ to $0$ with increasing ${{f_c}|\sin {\phi _m} - \sin \phi | + \frac{B}{2}|\sin \phi |}$ from $0$ to $2f_c/N$. First, we consider a beamforming codebook with an odd number of beams (or $L$) as shown in Fig. \ref{wor}. It can be verified that the maximum of (\ref{61}) is achieved at $\phi_m =\phi=0$ and the wideband beam gain is  $N$. On the other hand, its minimum occurs at $\phi=\pm \pi/2$ and $\phi_m=\phi_1$ or $\phi_L$, with $|\sin {\phi _m} - \sin \phi |=1/L$. In this case, the wideband beam gain is given by
\begin{equation}\label{worsg}
G(\phi ,{\bf{a}}({\phi _m})) = \left| {\frac{{\sin \left[ {N\pi (2{f_c} + BL)/4{f_c}L} \right]}}{{\sqrt{N}\sin \left[ {\pi (2{f_c} + BL)/4{f_c}L} \right]}}} \right|^2.
\end{equation}
As $N\pi\left( {2{f_c} + BL} \right)/4{f_c}L$ increases from $0$ to $\pi$, the wideband beam gain decreases from the maximum to $0$. Thus, to achieve the positive wideband gain by the narrowband beamforming codebook given in (\ref{codebook}), a necessary  condition is 
\begin{equation}
\begin{split}
N\pi \left( {2{f_c} + BL} \right)/4{f_c}L < \pi \Rightarrow N < \frac{{4{f_c}L}}{{2{f_c} + BL}} < \frac{{4{f_c}}}{B} = \frac{4}{\kappa }
\end{split}.
\end{equation}
While for the codebook with an even number of beams, it can be shown that the maximum of (\ref{61}) is still achieved at $\phi=0$ but cannot reach $N$ due to the beam misalignment. The minimum of (\ref{61})  is still given by (\ref{worsg}), which thus completes the proof.

\section{Derivation of  (\ref{close})  }
By substituting ${\bf{w}} = {\bf{a}}({\phi _m})$ into (\ref{ggain}), the beam gain at frequency $f$ in the direction of $\phi$ is given by

\begin{align}\label{beamgain}
g(f,\phi ,{\bf{w}}) &= \left|\frac{1}{\sqrt{N}} {\sum\limits_{n = 1}^N {{e^{j(n - 1)\pi [\sin {\phi _m} - (1 + \frac{f}{{{f_c}}})\sin \phi ]}}} } \right|^2 \notag\\
&= \left| {\frac{{\sin \frac{{N\pi [{f_c}\sin {\phi _m} - (f + {f_c})\sin \phi ]}}{{2{f_c}}}}}{{\sqrt{N}\sin \frac{{\pi [{f_c}\sin {\phi _m} - (f + {f_c})\sin \phi ]}}{{2{f_c}}}}}} \right|^2  \\
 &= \left| {\frac{{\sin \frac{{N\pi }}{{2{f_c}}}\left[ {{f_c}(\sin {\phi _m} - \sin \phi ) - f\sin \phi } \right]}}{{\sqrt{N}\sin \frac{\pi }{{2{f_c}}}\left[ {{f_c}(\sin {\phi _m} - \sin \phi ) - f\sin \phi } \right]}}} \right|^2,\notag
\end{align}

where the second equation is due to the following relation,
\begin{align}\label{68}
\sum\limits_{n = 1}^N {e^{j(n - 1)x}} &=  \frac{{1 \!-\! {e^{jNx}}}}{{1 \!-\! {e^{jx}}}} \!=\!  {\frac{{{e^{j\frac{{Nx}}{2}}}({e^{j\frac{{Nx}}{2}}} \!-\! {e^{ - j\frac{{Nx}}{2}}})}}{{{e^{j\frac{x}{2}}}({e^{j\frac{x}{2}}} \!-\! {e^{ - j\frac{x}{2}}})}}} \notag\\
& = {e^{j\frac{{(N - 1)x}}{2}}}\left[ {{{\sin \frac{{Nx}}{2}} \mathord{\left/
 {\vphantom {{\sin \frac{{Nx}}{2}} {\sin \frac{x}{2}}}} \right.
 \kern-\nulldelimiterspace} {\sin \frac{x}{2}}}} \right].
\end{align}
To study the property of $g(f,\phi ,{\bf{w}})$, we define a function $D(x) = \sin \left( {N\pi x} \right)/\sin \left( {\pi x} \right)$. With the increase of $|x|$ from $0$ to $1/N$, it can be verified that $D(x) $ decreases monotonically from $N$ to $0$. Thus, considering that $|{{f_c}(\sin {\phi _m} - \sin \phi ) - f\sin \phi }| \leq 2f_c/N$ for $f \in [ - \frac{B}{2},\frac{B}{2}]$, $g(f,\phi ,{\bf{w}})$ should decrease with $|{{f_c}(\sin {\phi _m} - \sin \phi ) - f\sin \phi }|$. Then, in the case of $\phi_m>\phi$, the minimum of $g(f,\phi ,{\bf{w}})$ is achieved at
\begin{equation}
f = \left\{ {\begin{aligned}
&{ - B/2,\;\;{\rm{  for }}\;\phi  \ge 0,{\rm{ }}}\\
&\;{B/2,\;\;\;\;{\rm{   for }}\;\phi  < 0}.
\end{aligned}} \right.
\end{equation}
In the case of $\phi_m<\phi$, the minimum of  $g(f,\phi ,{\bf{w}})$  is achieved at 
\begin{equation}
f = \left\{ {\begin{aligned}
&{ \;B/2,\;\;\;\;{\rm{  for }}\;\phi  \ge 0,{\rm{ }}}\\
&{-B/2,\;\;{\rm{   for }}\;\phi  < 0}.
\end{aligned}} \right.
\end{equation}
Finally, in the case of  $\phi_m=\phi$, the minimum  of $g(f,\phi ,{\bf{w}})$ is achieved at $f=\pm B/2$.  Thus, we obtain 
\begin{align}
&\mathop {\min }\limits_{f \in [ - \frac{B}{2},\frac{B}{2}]} g(f,\phi ,{\bf{w}}) =\\
& \left| {\frac{{\sin \frac{{N\pi }}{{2{f_c}}}\left[ {{f_c}|\sin {\phi _m} - \sin \phi | + \frac{B}{2}|\sin \phi |} \right]}}{{\sqrt{N}\sin \frac{\pi }{{2{f_c}}}\left[ {{f_c}|\sin {\phi _m} - \sin \phi | + \frac{B}{2}|\sin \phi |} \right]}}} \right|^2.
\end{align}
\section{Proof of Proposition 2}
Let $v=\frac{{2{f_c} + BL}}{{4{f_c}L}}\pi$. The worst-case performance can be rewritten as 
\begin{equation}\label{m68}
{\Gamma _{{\rm{worst}}}}({\cal W}^{\rm{nar}})= \frac{{{{\sin }^2}(Nv)}}{{N\sin v}}.
\end{equation}
${\Gamma _{{\rm{worst}}}}({\cal W}^{\rm{nar}})$ is a concave function with respect to $N$ for $N \in [1, {\frac{{4{f_c}L}}{{(2{f_c} + BL)}}})$. The first order derivative of (\ref{m68}) with respect to $N$ is given by 
\[\frac{{2vN\sin (Nv)\cos (Nv)\sin v - \sin v  {{\sin }^2}(Nv)}}{{{N^2}{{\sin }^2}v}}.\]
Considering that $\frac{{2{f_c} + BL}}{{4{f_c}L}} < 1$ and $N \in [1, {\frac{{4{f_c}L}}{{(2{f_c} + BL)}}})$, we have $\sin v > 0$ and $\sin (Nv) > 0$.
Then, the optimal condition can be simplified as 
\begin{equation}
2vN\cos (Nv) - \sin (Nv) = 0.
\end{equation}
Let $x=Nv$. It is then equivalent to finding the intersection of $2x$ and $\tan x$ for $x\in(0,\pi]$. Based on numerical search, the solution is given by $x=1.166$. Thus, we have 
\begin{equation}
N = \frac{{1.166}}{v} = \frac{{1.166 \times 4{f_c}L}}{{(2{f_c} + BL)\pi }} = \frac{{1.485{f_c}L}}{{2{f_c} + BL}}.
\end{equation}
Since $N$ is an integer, we finally  obtain (\ref{optinn}).
\section{Proof of Lemma 1}
The optimal solution to $({\rm{P2}}{\text{-}}p)$ is ${\bf{w}}_p^{\rm{opt}}$. Under the condition that  $\Omega_+ ^p-\Omega_- ^p=\Omega_+ ^q-\Omega_- ^q$, the problem $({\rm{P2}}{\text{-}}q)$ can be rewritten as 
\begin{equation*}
\begin{aligned}
({\rm{P2}}{\text{-}}q):\;\;\Gamma _{\rm{worst}}^q=& \mathop {\max }\limits_{{\bf{w}}_q} \mathop {\min }\limits_{\hat f}   \left| {{\bf{h}}{{(\hat f )}^H}{\bf{w}}_q} \right|,\\
&{\rm{s}}.{\rm{t}}.\;{\bf{h}}(\hat f ) = {\left[ {1,{e^{  j\pi\hat f }},...,{e^{  j\pi(N - 1)\hat f  }}} \right]^T},\;\;\\
&\;\;\;\;\;\;\hat f \in [\Omega_- ^p+T,\Omega_+ ^p+T],\\
&\;\;\;\;\;\;|{\bf{w}}_q(i)| = 1,\;i = 1,...,N,
\end{aligned}
\end{equation*}
where $T=\Omega_- ^q-\Omega_- ^p=\Omega_+ ^q-\Omega_+ ^p$. Then, it can be easily verified that $({\rm{P2}}{\text{-}}q)$ is equivalent to
\begin{equation*}
\begin{aligned}
({\rm{P2}}{\text{-}}q):\;\;\Gamma _{\rm{worst}}^q &= \mathop {\max }\limits_{{\bf{w}}_q} \mathop {\min }\limits_{\hat f}   \left| \big[{{\bf{h}}{{(\hat f )}\odot{\bf{h}}{(T)}\big] ^H}{\bf{w}}_q} \right|,\\
&{\rm{s}}.{\rm{t}}.\;{\bf{h}}(\hat f ) = {\left[ {1,{e^{ j\pi\hat f }},...,{e^{  j\pi(N - 1)\hat f  }}} \right]^T},\;\;\\
&\;\;\;\;\hat f \in [\Omega_- ^p,\Omega_+ ^p],\;|{\bf{w}}_q(i)| = 1,\;i = 1,...,N.
\end{aligned}
\end{equation*}
 By regarding ${\bf{h}}{(T)^\dag } \odot {\bf{w}}$ as the effective variable vector, where $\dag$ is the conjugate operator,  the problem is the same as $({\rm{P2}}{\text{-}}p)$. Thus, the optimal solution is ${\bf{w}}_p^{\rm{opt}}$ and 
we have ${\bf{w}}_q^* = {\bf{w}}_p^{\rm{opt}} \odot {\bf{h}}(T)$.
\section{Proof of Proposition 3}
Let  ${\bf{w}}^{\rm{opt}}$ denote the optimal solution of ${\rm{P2}}{\text{-}}l)$. Then, the objective function of $(({\rm{P2}}{\text{-}}l)$ is upper-bounded by
\begin{align}\label{42}
\mathop {\max }\limits_{\bf{w}} \mathop {\min }\limits_{\hat f \in [\Omega_- ^l,\Omega_+ ^l]} \left| {{\bf{h}}{{(\hat f)}^H}{\bf{w}}} \right|^2 &= \mathop {\min }\limits_{\hat f \in [\Omega_- ^l,\Omega_+ ^l]}\left| {{\bf{h}}{{(\hat f)}^H}{{\bf{w}}^{\rm{opt}}}} \right|^2 \\
&\le \frac{1}{\Delta \Omega}\int_{\Omega_- ^l}^{\Omega_+ ^l} {{{\left| {{\bf{h}}{{(\hat f)}^H}{{\bf{w}}^{\rm{opt}}}} \right|}^2}d\hat f} ,\notag
\end{align}
where the inequality is due to the fact that the minimum value is no larger than the average value.  For practical system parameters, $\Delta \Omega$ is much smaller than 1. Since ${\bf{h}}{(\hat f,\phi )^H}{{\bf{w}}^{\rm{opt}}} = \sum\nolimits_{n = 1}^N {{{\bf{w}}^{\rm{opt}}}(n){e^{j\pi (n - 1)\hat f}}} $ is a periodic function with a period of $2$,  (\ref{42}) can be further upper-bounded by
\begin{align}\label{45}
&\int_{\Omega_- ^l}^{\Omega_+^l} {{{\left| {{\bf{h}}{{(\hat f)}^H}{{\bf{w}}^{\rm{opt}}}} \right|}^2}d\hat f}  \le \int_{ -1}^{1} {{{\left| {\sum\limits_{n = 1}^N {{{\bf{w}}^{\rm{opt}}}(n){e^{j\pi (n - 1)\hat f}}} } \right|}^2}d\hat f} \notag \\
& = \int_{ -1}^{1} {\sum\limits_{n = 1}^N {\sum\limits_{m = 1}^N {{{\bf{w}}^{\rm{opt}}}(n){{{\bf{\bar w}}}^{\rm{opt}}}(m){e^{j\pi (n - 1)\hat f}}} } {e^{ - j\pi (m - 1)\hat f}}d\hat f} \notag\\
 &= 2\sum\limits_{m = 1}^N {{{\bf{w}}^{\rm{opt}}}(n){{{\bf{\bar w}}}^{\rm{opt}}}(n)} \notag\\
 &\qquad \qquad  + \sum\limits_{n = 1}^N {\sum\limits_{m \ne n}^N {{{\bf{w}}^{\rm{opt}}}(n){{{\bf{\bar w}}}^{\rm{opt}}}(m)} } \int_{ -1}^{1} {{e^{j\pi (n - m)\hat f}}d\hat f} \notag \\
 &= 2\sum\limits_{m = 1}^N {{{\left| {{{\bf{w}}^{\rm{opt}}}(n)} \right|}^2}}  = 2.
\end{align}

By substituting (\ref{45}) into (\ref{42}), the proof is completed.

\section{Derivation of (\ref{thetaclose})  }
The beam response in the direction of the $i$-th intersection point can be written as 
 \begin{align}\label{78}
&{{{\bf{a}}}_N}{(\psi _{{\rm{itr}}}^i)^H}{{\bf{w}}_{\Delta \Omega }} = \sum\limits_{n = 1}^N {{{\bf{w}}_{\Delta \Omega }}(n)} {e^{j\pi (n - 1)\psi _{{\rm{itr}}}^i}}\\
&= {\sum\limits_{z = 1}^Z {\left[ {{e^{ - j\pi (z - 1){N_s}\psi _{{\rm{itr}}}^i}}{{{\bf{a}}}_{{N_s}}}(\psi _{{\rm{itr}}}^i)} \right]} ^H}\left[ {{e^{j{\theta _z}}}{{{\bf{a}}}_{{N_s}}}(\psi _{{\rm{pt}}}^z)} \right].\notag
 \end{align}
The left and right sides of the $i$-th intersection point are the $i$-th beam and the $(i+1)$-th beam, respectively. As the adjacent two beams have a dominant effect on the beam gain, we neglect the response of other beams and thus (\ref{78}) can be approximated as 
\begin{align}
 &{{{\bf{a}}}_N}{(\psi _{{\rm{itr}}}^i)^H}{{\bf{w}}_{\Delta \Omega }} \\
 &\approx {e^{j\pi (i - 1){N_s}\psi _{{\rm{itr}}}^i}}{e^{j{\theta _i}}}{{{\bf{a}}}_{{N_s}}}{(\psi _{{\rm{itr}}}^i)^H}{{{\bf{a}}}_{{N_s}}}(\psi _{{\rm{pt}}}^i) \notag\\
&\qquad \qquad + {e^{j\pi i{N_s}\psi _{{\rm{itr}}}^i}}{e^{j{\theta _{i + 1}}}}{{{\bf{a}}}_{{N_s}}}{(\psi _{{\rm{itr}}}^i)^H}{{{\bf{a}}}_{{N_s}}}(\psi _{{\rm{pt}}}^{i + 1})\notag\\
 &= {e^{j\left[ {\pi (i - 1){N_s}\psi _{{\rm{itr}}}^i + {\theta _i}} \right]}} \times \notag\\
 & \left[ {\underbrace {\sum\limits_{n = 1}^{{N_s}} {{e^{j\frac{{\pi (n - 1)\Delta \Omega }}{{2Z}}}}} }_{{\rm{term}}1} + \underbrace {{e^{j\pi {N_s}\psi _{{\rm{itr}}}^i}}{e^{j({\theta _{i + 1}} - {\theta _i})}}\sum\limits_{n = 1}^{{N_s}} {{e^{ - j\frac{{\pi (n - 1)\Delta \Omega }}{{2Z}}}}} }_{{\rm{term}}2}} \right].\notag
\end{align}

To maximize the beam gain, we should make the two above terms in-phase by properly setting $\theta_{i}$ and $\theta_{i+1}$. According to (\ref{68}), the phase of the first term is $\frac{{\pi ({N_s} - 1)\Delta \Omega }}{{4Z}}$. Thus,  the following relationship should be satisfied:

\begin{align}
\frac{{\pi ({N_s} - 1)\Delta \Omega }}{{4Z}} &= \pi {N_s}\psi _{{\rm{itr}}}^i + ({\theta _{i + 1}} - {\theta _i}) - \frac{{\pi ({N_s} - 1)\Delta \Omega }}{{4Z}} \notag\\
 \Rightarrow {\theta _{i + 1}} - {\theta _i} = &\frac{{\pi ({N_s} - 1)\Delta \Omega }}{{2Z}} - \pi {N_s}\psi _{{\rm{itr}}}^i \notag\\
=&\frac{{\pi ({N_s} - 1)\Delta \Omega }}{{2Z}} - \pi {N_s}\left( {\frac{{i\Delta \Omega }}{Z} - \frac{{\Delta \Omega }}{2}} \right)\notag\\
=& \left[ {\frac{{({N_s} - 1)}}{{2Z}} + \frac{{{N_s}}}{2}} \right]\pi \Delta \Omega  - \frac{{\pi {N_s}\Delta \Omega }}{Z}i \notag\\
 \buildrel \Delta \over =& a - bi.
\end{align}

To maximize the beam gain for all intersection points, the above relationship should hold for $i=1,2,...,Z-1$. By letting $\theta_1=0$, we can obtain $\{\theta _ z\}_{z=1}^Z$ as 
\begin{align}
{\theta _z} &= {\theta _z} - {\theta _{z - 1}} + {\theta _{z - 1}} - {\theta _{z - 2}} + ... + {\theta _2} - {\theta _1} \notag\\
&= a - b(z - 1) + a - b(z - 2) + ... + a - b \notag\\
 &= a(z - 1) - b\frac{{z(z - 1)}}{2} \notag\\
 &= \left[ {\frac{{({N_s} - 1)}}{{2Z}} + \frac{{{N_s}}}{2}} \right]\pi \Delta \Omega (z - 1) - \frac{{\pi {N_s}\Delta \Omega }}{{2Z}}z(z - 1)\notag\\
 &= \left[ {\frac{{(1 + Z){N_s} - 1}}{{2Z}}(z - 1) - \frac{{{N_s}}}{{2Z}}z(z - 1)} \right]\pi \Delta \Omega \notag\\
 &= \frac{{(Z - z + 1){N_s} - 1}}{{2Z}}(z - 1)\pi \Delta \Omega, \;\; \forall z.
\end{align}
\end{appendices}

\end{document}